\documentclass[useAMS,usenatbib]{mn2e}

\usepackage[english]{babel}
\usepackage[ansinew]{inputenc}
\usepackage{amssymb,amsmath}
\usepackage{graphicx} 
\usepackage{natbib} 
\usepackage{wrapfig}
\usepackage{color}
\usepackage{ulem}

\newcommand{\nmax}{N_\textrm{max}}

\newcommand{\diag}{\textrm{diag}}

\title[Quantifying galaxy shapes: S\'ersiclets and beyond]{Quantifying galaxy shapes: S\'ersiclets and beyond}

\author[R. Andrae, P. Melchior \& K. Jahnke (2011)]{Ren\'e Andrae$^{1}$\thanks{E-mail:
andrae@mpia-hd.mpg.de}, Peter Melchior$^{2,3,4}$ and Knud Jahnke$^{1}$\\
$^{1}$Max-Planck-Institut f\"ur Astronomie, K\"onigstuhl 17, 69117 Heidelberg, Germany\\
$^{2}$Institut f\"ur Theoretische Astrophysik, Zentrum f\"ur Astronomie, Albert-Ueberle-Str.\ 2, 69120 Heidelberg, Germany\\
$^{3}$Center for Cosmology and Astro-Particle Physics, The Ohio State University, 191 W.\ Woodruff Ave., Columbus, Ohio 43210, USA\\
$^{4}$Department of Physics, The Ohio State University, 191 W.\ Woodruff Ave., Columbus, Ohio 43210, USA}

\begin{document}

\date{Accepted 2011 June 28. Received 2011 February 22.}

\pagerange{\pageref{firstpage}--\pageref{lastpage}} \pubyear{2010}

\maketitle

\label{firstpage}

\begin{abstract}
Parametrisation of galaxy morphologies is a challenging task, for instances in shear measurements of weak gravitational lensing or investigations of formation and evolution of galaxies. The huge variety of different morphologies requires a parametrisation scheme that is highly flexible and that accounts for certain morphological observables, such as ellipticity, steepness of the radial light profile and azimuthal structure. In this article, we revisit the method of s\'ersiclets, where galaxy morphologies are decomposed into a set of polar basis functions that are based on the S\'ersic profile. This approach is justified by the fact that the S\'ersic profile is the first-order Taylor expansion of any real light profile. We show that s\'ersiclets indeed overcome the modelling failures of shapelets in the case of early-type galaxies. However, s\'ersiclets implicate an unphysical relation between the steepness of the light profile and the spatial scale of the polynomial oscillations, which is not necessarily obeyed by real galaxy morphologies and can therefore give rise to modelling failures. Moreover, we demonstrate that s\'ersiclets are prone to undersampling, which restricts s\'ersiclet modelling to highly resolved galaxy images. Analysing data from the weak-lensing \textsc{Great08} challenge, we demonstrate that s\'ersiclets should not be used in weak-lensing studies. We conclude that although the s\'ersiclet approach appears very promising at first glance, it suffers from conceptual and practical problems that severly limit its usefulness. In particular, s\'ersiclets do not provide high precision results in weak-lensing studies. Finally, we show that the S\'ersic profile can be enhanced by higher-order terms in the Taylor expansion, which can drastically improve model reconstructions of galaxy images. When orthonormalised, these higher-order profiles can overcome the problems of s\'ersiclets while preserving their mathematical justification. However, this method is computationally expensive.
\end{abstract}



\section{Introduction}

There are two scientific key diagnostics in modern investigations of galaxy formation and evolution, namely star-formation rates and galaxy morphologies. Galaxy morphologies are known to correlate to different degrees with other more physical parameters such as star formation \citep[e.g.,][]{Kennicutt1998} or environment \citep[e.g.,][]{Wel2010}. Although morphology by itself is not a fundamental parameter of a galaxy, it provides a direct observable. Conversely, e.g., star-formation rates are derived quantities based on additional assumptions and extrapolations \citep[e.g.,][]{Rosa2002}. Studies of galaxy morphologies are therefore an important complementary means for studying the physics of galaxy formation. Furthermore, certain investigations are solely based on morphologies, e.g., weak-lensing measurements \citep[e.g.,][]{Bernstein2002} or studies of angular-momentum alignments of spiral galaxies \citep[e.g.,][]{Slosar2009,Andrae2011}. Concerning weak gravitational lensing, the investigation of shape-measurement algorithms currently is a very active and important field of research since established and well known procedures fail to measure object shapes accurately enough to fully exploit the potential of future large-scale imaging surveys \citep[e.g.][]{Bridle2010,Bernstein2010}. Concerning morphological classification, galaxies were traditionally described mostly by visual classifications. However, given the enormous amount of galaxies known from today's sky surveys, it is evident that an automated \textit{parametrisation} of galaxy morphologies is required. Consequently, there has been substantial effort to define automated parametrisation schemes for galaxy morphologies \citep[e.g.,][to name just a few]{Sersic1968,Abraham1996,Abraham2003,Bershady2000,Conselice2003,Lotz2004,Lotz2008}. Unfortunately, these parametrisation schemes usually invoke rather restrictive assumptions, such that they lack the flexibility to describe the huge variety of different galaxy morphologies present in modern databases \citep{Andrae2010b}. The S\'ersic profile \citep{Sersic1968} is one example in this respect: The S\'ersic model is the first-order Taylor expansion of any real radial light profile (cf.\ Appendix \ref{app:1st_order_Taylor}) and therefore provides a good first-order approximation to real galaxies. However, second-order effects -- e.g., deviations of the radial profile from the S\'ersic form or the appearance of azimuthal structures such as star-forming regions, galactic bars or spiral-arm patterns -- are typically not negligible, and hence there have been several attempts to enhance the S\'ersic profile:
\begin{itemize}
\item Mixtures of two or more S\'ersic profiles \citep[e.g.][]{Simmat2010}.
\item Modification of the S\'ersic profile by Fourier modes in order to describe azimuthal structures \citep[e.g., \textsc{Galfit} 3,][]{Peng2010}.
\item Orthogonalisation of the S\'ersic profile in order to describe azimuthal structures \citep{Ngan2008}.
\end{itemize}
The idea of using basis functions in order to parametrise galaxy morphologies is not new. For instances, ``shapelets'' \citep{Refregier2003} are a set of basis functions based on the Gaussian profile. The appealing properties of shapelets gave rise to a number of investigations of galaxy morphologies \citep[e.g.,][]{Kelly2004,Kelly2005,Andrae2010a} and weak-lensing studies \citep[e.g.,][]{Massey2007,Ferry2008}. However, as was recently shown by \citet{Melchior2009b}, shapelets can suffer from strong biases that originate from the radial light profiles of galaxies potentially being much steeper than a Gaussian profile, in particular for early-type galaxies. These biases manifest themselves as ring-like artefacts in the models and the residual maps (cf.~Sect.~\ref{sect:image_decompositions}). They are able to wash out virtually all the information about ellipticity of an object and are likely to affect other morphological quantities in a similar way. Hence, it was obvious to introduce a set of basis functions based on the S\'ersic profile \citep{Ngan2008}, called ``s\'ersiclets''. As these basis functions explicitly account for the steepness of radial profiles, they are highly flexible and seem to be promising candidates to provide a method for describing the huge variety of different galaxy morphologies.

This article is organised as follows. We define s\'ersiclet basis functions in Sect. \ref{sect:shapelets_and_sersiclets}. Section \ref{sect:systematic_tests} presents systematic tests of s\'ersiclets, working out their reliability and limitations. We investigate the performance of s\'ersiclets in an application to artificial weak-lensing data in Sect. \ref{sect:app:lensing}. In Sect.~\ref{sect:3rd_order_Taylor} we discuss the potential of enhancing the S\'ersic profile by higher-order Taylor expansions and briefly investigate the orthonormalisations of such light profiles. We conclude in Sect. \ref{sect:conclusions}. Further details are given in the appendices. Appendix~\ref{app:Sersic_and_Taylor} contains the proof that S\'ersic profiles are first-order Taylor expansions. Appendix~\ref{app:derivation_sersiclets} provides the analytic derivation of the s\'ersiclet basis functions. Finally, we explain in Appendix~\ref{sect:optimisation} how to fit s\'ersiclet models to imaging data.

\section{Polar s\'ersiclets}
\label{sect:shapelets_and_sersiclets}

We briefly comment on a prior attempt to introduce s\'ersiclets before we then give our definition.

\subsection{The first attempt to introduce s\'ersiclets}

\begin{figure}
\includegraphics[width=0.5\textwidth]{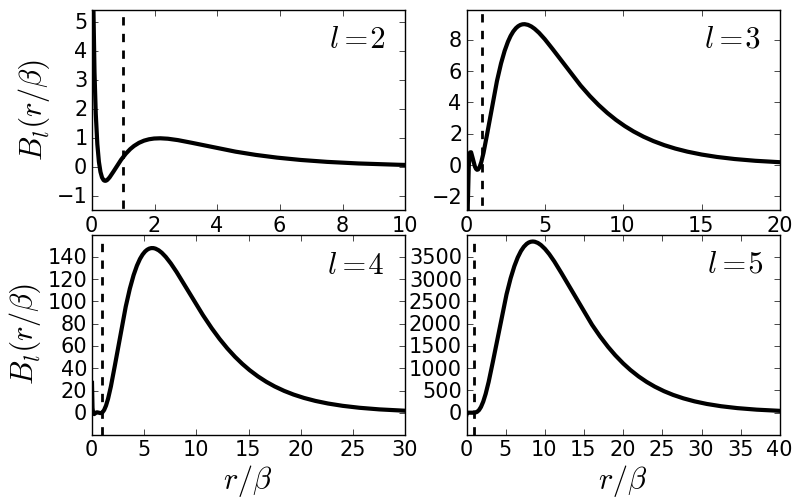}
\caption{Radial basis functions if orthonormalised on the integration interval $r\in [0,\beta]$ using a S\'ersic profile with index $n_S=2$ as weight function \citep{Ngan2008}. The vertical dashed lines indicate $r=\beta$, where the orthonormalisation interval ends. The S\'ersic profile is not able to suppress the polynomials outside the orthonormalisation interval.}
\label{fig:radial_basis_functions_true_sersiclets}
\end{figure}

Given the aforementioned problems of shapelets, it was obvious to orthonormalise the S\'ersic profile. Moreover, this choice is ``natural'' because the S\'ersic profile is the first-order Taylor expansion of any real light profile (cf.\ Appendix \ref{app:1st_order_Taylor}). This is likely to remove the strong biases of shapelets. The resulting basis functions are called s\'ersiclets and were first investigated by \citet{Ngan2008}.

However, \citet{Ngan2008} faced a severe problem with s\'ersiclets. In a simple test case, \citet{Ngan2008} observed that the polar s\'ersiclets were incapable of fitting a given object. Even when increasing the number of basis functions used for the decomposition, the model did not converge to the given image data. This directly implies that the basis functions constructed by \citet{Ngan2008} are not complete. We found out that this problem originates from the fact that \citet{Ngan2008} orthonormalised their basis functions out to one half-light radius -- not on the interval $r\in[0,\infty[$ -- but then extended their basis functions beyond this range. Figure \ref{fig:radial_basis_functions_true_sersiclets} shows what happens in this case. In the inner region, the basis functions may be orthonormal, but for larger radii the leading order monomial in the basis functions dominates over the S\'ersic profile. This creates an artificial bump outside the region of orthonormality, which appears in all radial modes of order $l>0$ and becomes the dominant feature. Consequently, these basis functions loose their linear independence, which explains the non-convergence observed by \citet{Ngan2008}.

In order to solve this problem, they suggested to discard all modes with azimuthal ``quantum number'' $m\neq 0$, i.e., all mode with azimuthal structure, and only maintain the radial modes. While this approach may be viable for weak-lensing applications, where azimuthal structures are rarely visible, it does not solve the actual problem, which is the extension beyond the orthonormalisation interval. Moreover, the resulting basis functions may still represent an expansion into radial modes, but it is not only radial but also azimuthal structure of galaxies that is interesting, e.g., in morphological classification. In fact, it is the ability to describe azimuthal structure in a (theoretically) well defined way that makes basis-function expansions of galaxy morphologies such a compelling approach. Consequently, \citet{JimenezTeja2011} criticised that this implementation of s\'ersiclets is incapable of describing irregular galaxies.

\subsection{Definition of polar s\'ersiclets}

\begin{figure}
\begin{minipage}{0.23\textwidth}
\includegraphics[width=1.0\textwidth]{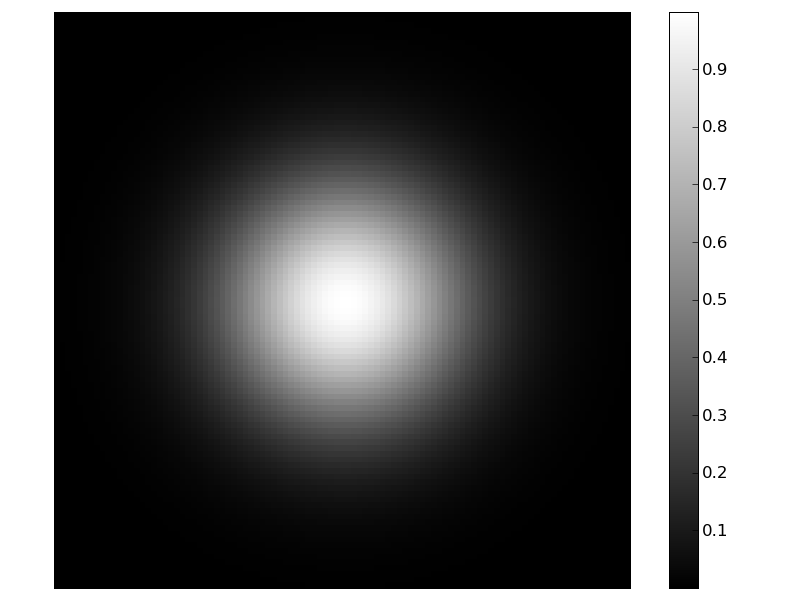}
\end{minipage}
\begin{minipage}{0.23\textwidth}
\includegraphics[width=1.0\textwidth]{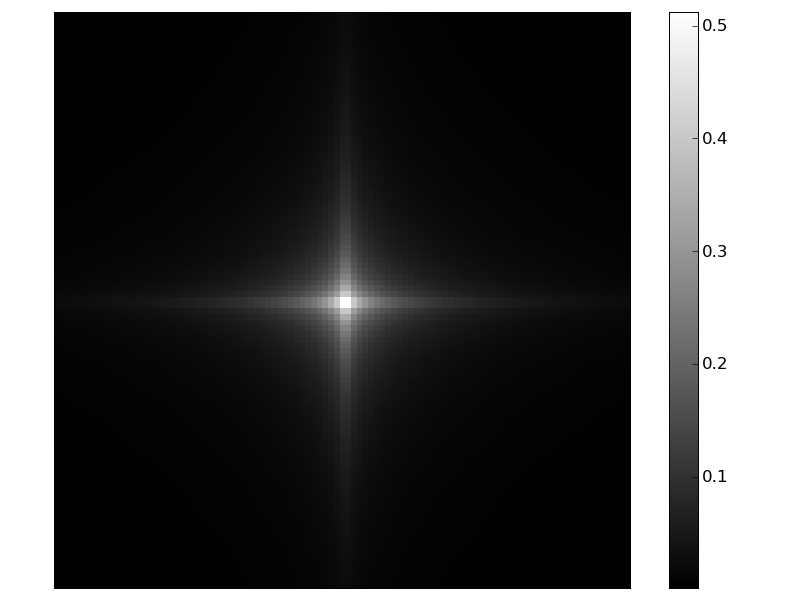}
\end{minipage}
\caption{Cartesian ground state using Gaussian profile (left panel) and S\'ersic profile with index $n_S=2$ (right panel). Unless we employ the Gaussian profile, the Cartesian weight function is not spherically symmetric.}
\label{fig:cartesian_groundstate}
\end{figure}

We now define the s\'ersiclet basis functions. In the case of shapelets, we can define both Cartesian and polar shapelets due to the mathematical features of the Gaussian weight function. However, for general S\'ersic profiles, we cannot meaningfully define Cartesian basis functions because the ground states hardly resemble any known galaxy morphology, as is shown in Fig.~\ref{fig:cartesian_groundstate}. Therefore, we have to introduce s\'ersiclets as polar basis functions.

\begin{figure}
\includegraphics[width=8cm]{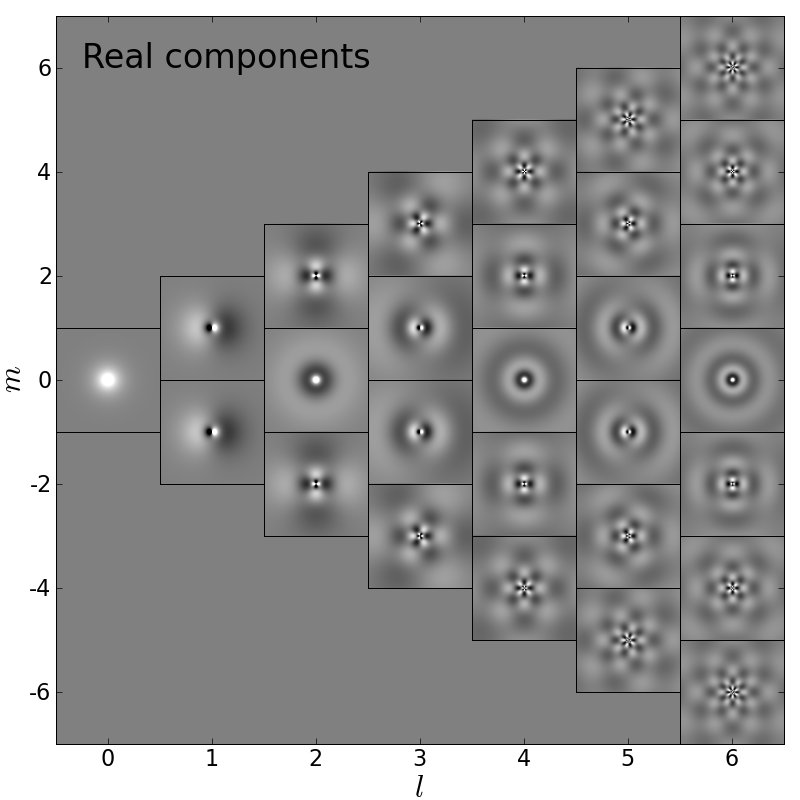}
\includegraphics[width=8cm]{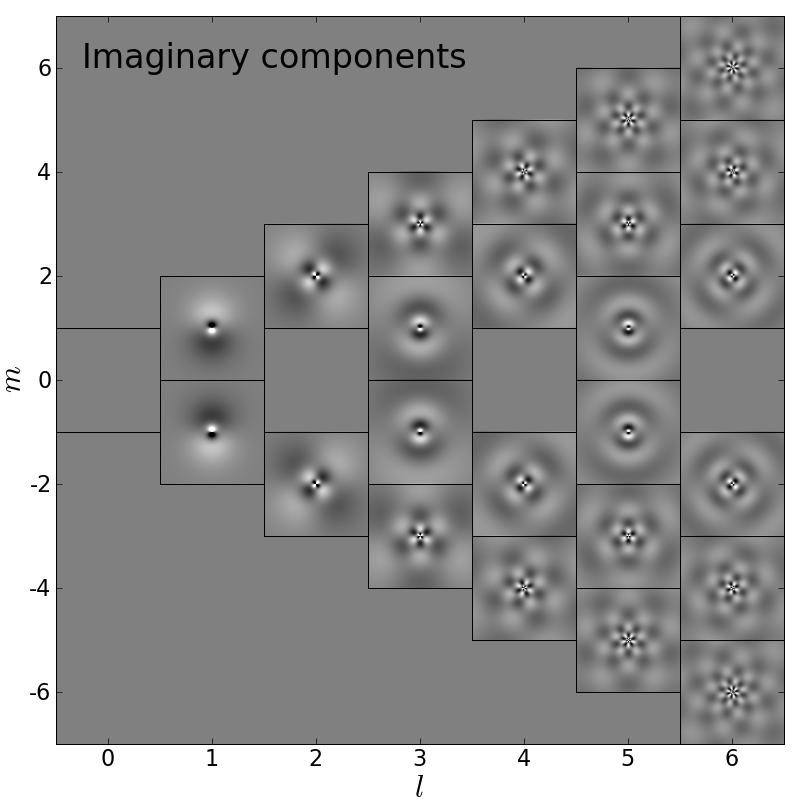}
\caption{Polar s\'ersiclet basis functions with $n_S=1$, $b=1$, and $\epsilon=0$. The real components of the complex-valued basis functions are shown in the top panel, the imaginary components in the bottom panel. The basis functions with $m=0$ are wholly real. We note that the polar basis functions exhibit lots of substructure in the central region. Moreover, these central substructures become very small with increasing radial order $l$.}
\label{fig:example_basis_functions}
\end{figure}

The crucial idea to overcome the problems faced by \citet{Ngan2008} is to realise that we \textit{cannot} expect a galaxy's structure being well captured within one half-light radius. For instances, disc galaxies may exhibit spiral-arm patterns in their outskirts. Therefore, we require orthonormality on the full interval $r\in[0,\infty[$. Apart from solving the problems reported by \citet{Ngan2008}, this also has the advantage that the orthonormal polynomials exist analytically. The s\'ersiclet basis functions are derived in Appendix~\ref{app:derivation_sersiclets} and are given by
\begin{equation}\label{eq:radial_basis_fct}
R_l(r) \propto L_l^{2n_S-1} \left[b\left(r/\beta\right)^{1/n_S}\right] \exp\left[-\frac{b}{2}\left(r/\beta\right)^{1/n_S}\right] \,\textrm{,}
\end{equation}
where the normalisation factor is also derived in Appendix~\ref{app:derivation_sersiclets}. Apart from the linear expansion coefficients, s\'ersiclets have the following nonlinear model parameters:
\begin{itemize}
\item The maximum order $N_\textrm{max}$.
\item The S\'ersic index $n_S$.
\item The scale radius $\beta$.
\item The centroid position $\vec x_0$, which enters via $r=|\vec x-\vec x_0|$.
\item The complex ellipticity $\epsilon$, which enters via a shear transformation \citep[e.g.][]{Bartelmann2001} thereby affecting $r$.\footnote{It is advantageous to use the real and imaginary parts of the complex ellipticity as model parameters, rather than orientation angle and axis ratio. Using the orientation angle as parameter would cause severe problems with convergence in the case of nearly spherical objects.}
\end{itemize}
The coefficient $b$ is defined by \citep{Graham2005}
\begin{equation}
\Gamma(2n_S)=2\gamma(2n_S,b) \,\textrm{,}
\end{equation}
where $\Gamma$ and $\gamma$ denote the complete and incomplete gamma functions. We explain in Appendix~\ref{sect:optimisation} how to fit these model parameters for given imaging data. In the case of $n_S=0.5$ and $b=1$ the polar s\'ersiclets reduce to the special case of polar shapelets.\footnote{Note that the polar functions defined by \citet{Massey2005} are not orthogonal. A correct functional form of polar shapelets is derived in Appendix~\ref{app:derivation_sersiclets}.} Furthermore, for $n_S=1$ and $b=1$, we get a set of basis function that could be called ``disclets'', since they have an exponential profile as weight function. Similarly, we could define ``deVaucouleurlets''. All these sets are special cases of Eq. (\ref{eq:radial_basis_fct}). Finally, we emphasise that when fitting a s\'ersiclet model, we always also fit a S\'ersic model, which is the ground state of the s\'ersiclets. Figure \ref{fig:example_basis_functions} displays an example of s\'ersiclet basis functions.

\subsection{Interpretation of the S\'ersic index}
\label{sect:interpret_nS}

In the case of s\'ersiclet basis functions, the S\'ersic index $n_S$ changes its interpretation. First, $n_S$ regulates the steepness of the weight function, which is a normal S\'ersic profile. Second, via the steepness of the weight function, $n_S$ also regulates the spatial scale on which the associated Laguerre polynomials oscillate. In simple words, there is a fixed relation between steepness of the weight function and oscillation scale of the polynomials. However, real morphologies do not necessarily obey such a relation. For instances, the steepness of the bulge is not related to the positions of spiral arms or star-forming regions. In practice, this can lead to modelling problems, if the steepness of the profile and the range of oscillation scales required for a faithful description of a given galaxy morphology do not match.

\section{Testing s\'ersiclets}
\label{sect:systematic_tests}

Before we can apply s\'ersiclets to scientific questions, we need to investigate the performance and fidelity of s\'ersiclets.

\subsection{Completeness}

\begin{figure}
\includegraphics[width=8cm]{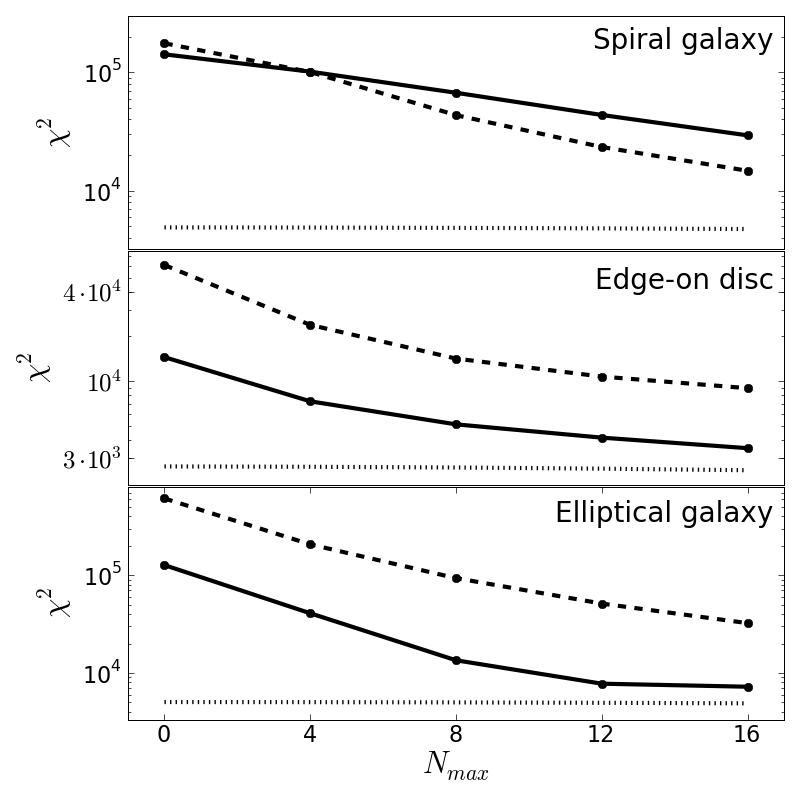}
\caption{Dependence of $\chi^2$ of s\'ersiclet decomposition (solid lines) and shapelet decompositions (dashed lines) on maximum order, $\nmax$, for the spiral galaxy, the edge-on disc, and the elliptical galaxy shown in Fig. \ref{fig:models_residuals}. For each example, we also show the number of pixels as a horizontal dotted line.}
\label{fig:examples_chi2_vs_Nmax}
\end{figure}

First, we need to test the completeness of s\'ersiclets in order to proof that we indeed overcome the problems reported by \citet{Ngan2008}. We decompose three real galaxy images from the Sloan Digital Sky Survey into s\'ersiclets with increasing maximum order. If our basis functions were not linearly independent, completeness would break down and the $\chi^2$-values would not decrease with increasing maximum order as in the case of \citet{Ngan2008}. Figure \ref{fig:examples_chi2_vs_Nmax} shows the results. First and foremost, the residuals of s\'ersiclets are decreasing with increasing maximum order, i.e., we indeed set up a set of basis functions that are linearly independent. Furthermore, it is interesting to compare the residuals of s\'ersiclets with those of circular shapelets. In the case of the spiral galaxy, s\'ersiclets yield residuals comparable to shapelets. This is not surprising, because shapelets excel in modelling extended objects with lots of substructure, such as a face-on spiral galaxy. However, in the case of the edge-on disc and the elliptical galaxy, the $\chi^2$-values of s\'ersiclets are substantially lower. Evidently, s\'ersiclets outperform  shapelets in modelling galaxies of these types, as expected from the steep light profiles.

\subsection{Image decompositions}
\label{sect:image_decompositions}

We have seen in the previous section that s\'ersiclet decompositions produce substantially lower residuals than (circular) shapelet decompositions when it comes to modelling galaxies that exhibit steep profiles, such as elliptical galaxies or edge-on discs. However, we still have to check whether s\'ersiclets indeed overcome the ring-like artefacts produced by shapelets.

\begin{figure*}
\includegraphics[width=16cm]{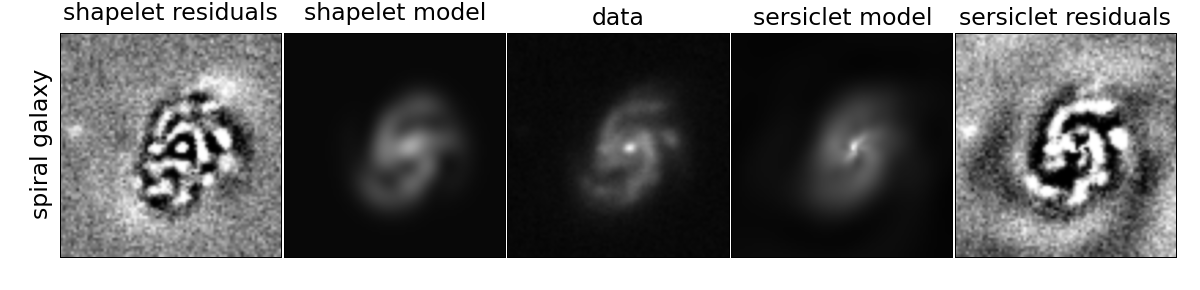}
\includegraphics[width=16cm]{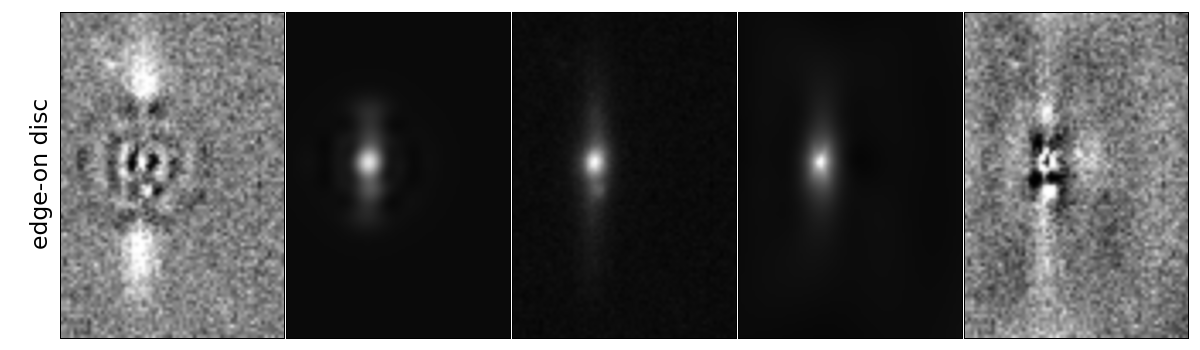}
\includegraphics[width=16cm]{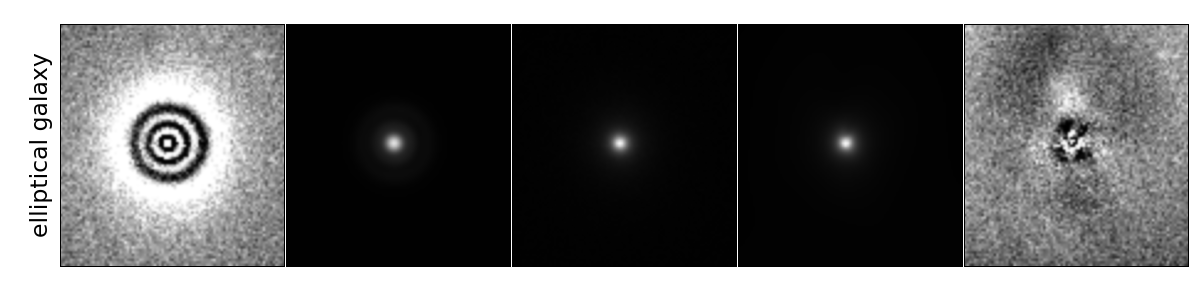}
\caption{Models and residual maps resulting from s\'ersiclets and shapelets for the spiral galaxy (top row), the edge-on disc (central row), and the elliptical galaxy (bottom row). All models used $\nmax=8$. For the sake of visualisation, the colour code of the model and data maps is nonlinear and ranges from $-4.5\sigma$ to the maximum value. The colour code of the residual maps is linear and ranges from $-4.5\sigma$ to $+4.5\sigma$, where $\sigma$ denotes the standard deviation of the background noise. The \textit{peak} significance of the imaging data is $132\sigma$ for the spiral galaxy, $101\sigma$ for the edge-on disc, and $610\sigma$ for the elliptical galaxy.}
\label{fig:models_residuals}
\end{figure*}

Figure \ref{fig:models_residuals} compares the best-fitting models and residuals of (circular) shapelets and (elliptical) s\'ersiclets using $\nmax=8$ for the three test galaxies. As expected from the similar $\chi^2$-values, in the case of the spiral galaxy, both shapelets and s\'ersiclets perform well in modelling the spiral-arm patterns. S\'ersiclets fit the central region better than shapelets, whereas shapelets tend to describe the outskirts better. As expected from a s\'ersiclet model with $n_S>0.5$, the polynomial oscillations appear on smaller scales than for the shapelet model with $n_S=0.5$. In the case of the edge-on disc, s\'ersiclets are clearly superior. First, the ring-like artefacts of shapelets are gone. Second, the intrinsic ellipticity of the s\'ersiclet model allows the basis functions to also describe the outermost regions of the disc, whereas these regions go almost unfitted by circular shapelets. In the case of the elliptical galaxy, the ring-like artefacts are very prominent in the shapelet residuals. Conversely, the s\'ersiclet residuals do not exhibit any artefacts of this kind, i.e., the steepness of the light profile is indeed described properly.

\subsection{Orthonormality and sampling\label{sect:orthogonality_test}}

\begin{figure}
\includegraphics[width=84mm]{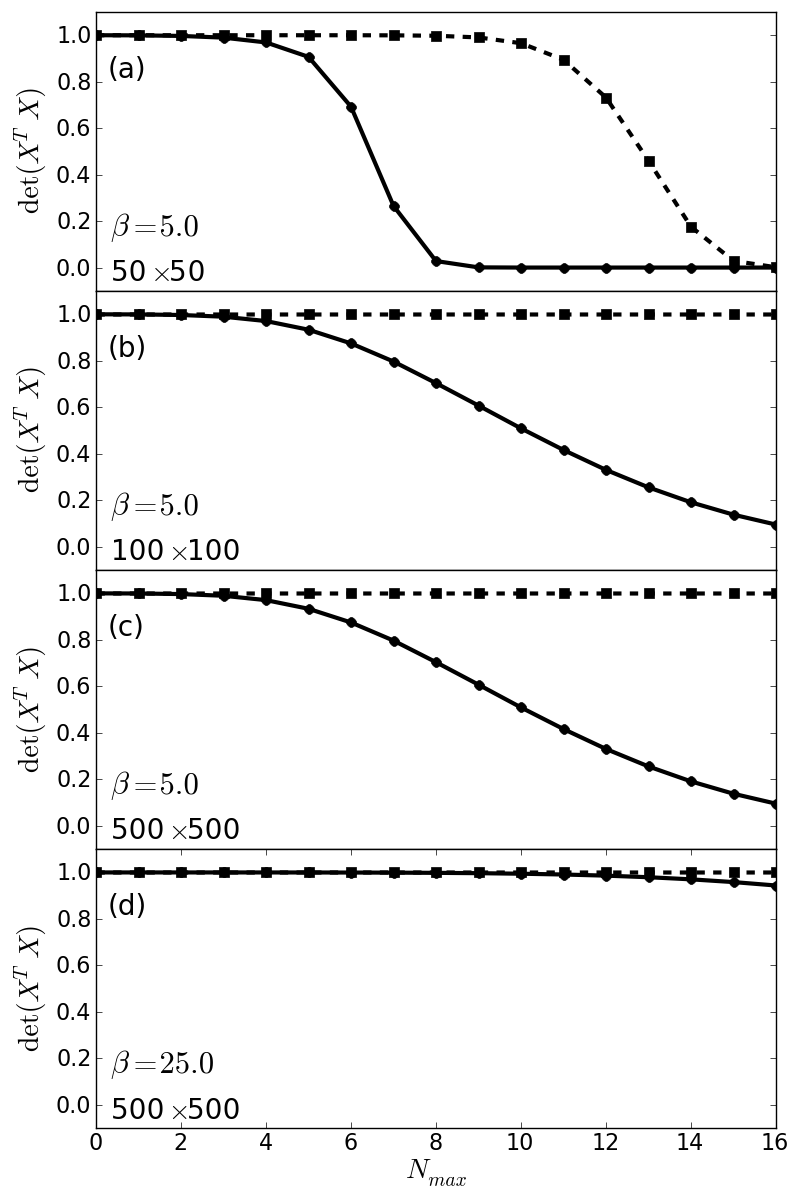}
\caption{Determinant of $X^T\cdot X$ for Cartesian (dashed lines with squares) and polar (solid lines with dots) shapelets for increasing maximum order $\nmax$, different scale radii $\beta$ and grid sizes. Panel a: At large $\nmax$ both Cartesian and polar shapelets deviate from $\det(X^T\cdot X)=1$. Panel b: Cartesian shapelets maintain $\det(X^T\cdot X)=1$ for all $\nmax$, proving that boundary effects are now negligible. Panel c: Polar shapelets still differ from $\det(X^T\cdot X)=1$ at large $\nmax$ on this enlarged grid, i.e., this is not a boundary effect. Panel d: Polar shapelets now also maintain $\det(X^T\cdot X)=1$, i.e., the remaining effect was due to undersampling.}
\label{fig:det_XtX_shapelets}
\end{figure}

Let $\Sigma$ denote the pixel covariance matrix of the noise in the imaging data and $X$ the design matrix (see Appendix~\ref{app:ML_Solution}). In order to get the maximum-likelihood estimate of the expansion coefficients from Eq.\ (\ref{eq:ML_estimate_coeffs}), the matrix $X^T\cdot\Sigma^{-1}\cdot X$ -- or $X^T\cdot X$ for uncorrelated pixel noise -- needs to be invertible. If it were not for pixellation, finite image grids, and the PSF, $X^T\cdot X$ should be an identity matrix due to the orthonormality of the basis functions. We now investigate the orthonormality of s\'ersiclets in two tests.

First, we compare the orthonormality of \textit{polar} shapelets (s\'ersiclets with $n_S=0.5$ and $b=1$) and \textit{Cartesian} shapelets of \citet{Melchior2007}. We sample polar and Cartesian shapelets of constant scale radius $\beta=5$ on pixel grids of sizes 50$\times$50, 100$\times$100, and 500$\times$500 and compute the determinant of $X^T\cdot X$ for maximum orders $0\leq \nmax \leq 16$, where $\det(X^T\cdot X)<1$ indicates a violation of orthonormality. Figure \ref{fig:det_XtX_shapelets} shows the results of this test. In the case of the 50$\times$50 grid (panel (a)) both Cartesian and polar shapelets suffer from non-orthonormality for increasing $\nmax$.\footnote{In particular, for polar shapelets and $\nmax\geq 8$ the determinant of $X^T\cdot X$ is zero, i.e., $X^T\cdot X$ is not invertible and the estimator of the expansion coefficients (Eq.\ (\ref{eq:ML_estimate_coeffs})) breaks down.} This problem can be caused either by boundary truncation or undersampling of higher-order modes (with rapidly oscillating polynomials). Therefore, in panel (b) we increase the pixel grid from 50$\times$50 to 100$\times$100. The non-orthonormality of Cartesian shapelets is now cured, i.e., it has indeed been caused by boundary truncation. However, polar shapelets still exhibit non-orthonormality. Therefore, in panel (c) we increase the pixel grid again now from 100$\times$100 to 500$\times$500. The behaviour of polar shapelets is unchanged, i.e., the non-orthonormality is not caused by boundary trunction. In order to demonstrate that the non-orthonormality is caused by undersampling, we increase in panel (d) the object size from $\beta=5$ to $\beta=25$ while keeping the 500$\times$500 pixel grid. The non-orthonormality of polar shapelets is now almost cured, which verifies that this was an undersampling effect. Evidently, undersampling has a larger impact on polar shapelets than on Cartesian shapelets. The reason is that polar basis functions exhibit most of their structure in the central region (cf. Fig. \ref{fig:example_basis_functions}) and therefore suffer strongly from pixellation. Loosely speaking, \textit{polar} basis functions do not appreciate being sampled on a \textit{Cartesian} pixel grid.

\begin{figure*}
\includegraphics[width=160mm]{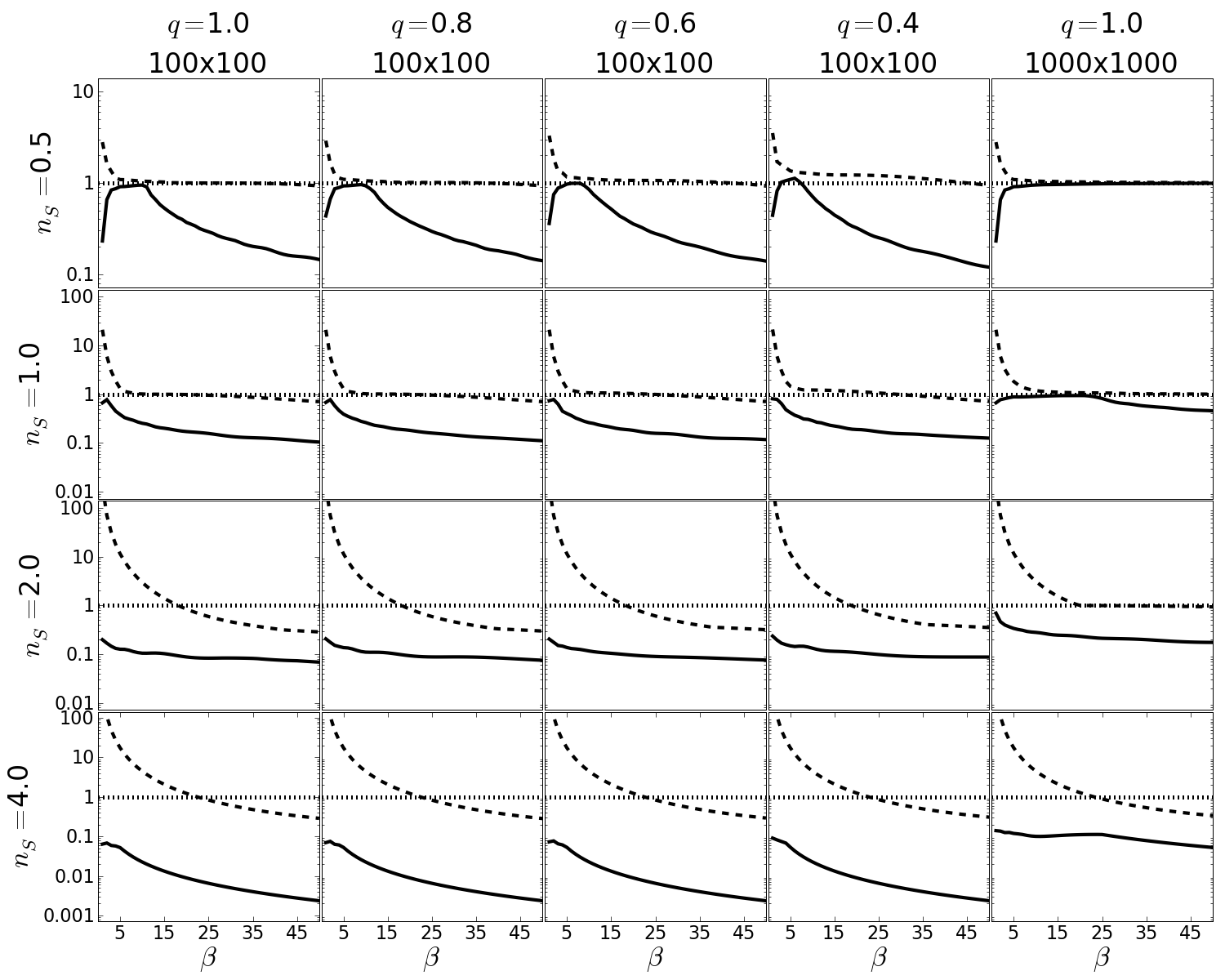}
\caption{Smallest (solid line) and largest (dashed line) diagonal elements of matrix $X^T\cdot X$ as a function of $\beta$ at $\nmax=12$. Deviation from unity (dotted line) indicates the loss of orthonormality. The image size was 100$\times$100 pixels (left columns) and 1000$\times$1000 pixels (right column). We used $b=2\log 3$, such that at one scale radius the weight function drops to one third of its central value.}
\label{fig:ONtest_smallest_largest_diag_elem}
\end{figure*}

Second, having attested problems of polar shapelets with orthonormality due to undersampling, we now investigate the orthornomality of (polar) s\'ersiclets in general. We compare the value of the largest and smallest diagonal element of $X^T\cdot X$, respectively, repeating the same test as \citet{Berry2004} did for shapelets. We evaluate s\'ersiclet models of maximum order $\nmax=12$, S\'ersic indices $n_S=0.5,1,2,4$ and axis ratios $q=1,0.8,0.6,0.4$ on a 100$\times$100 pixel grid for varying scale radii $\beta$. There is no PSF in this test. Figure \ref{fig:ONtest_smallest_largest_diag_elem} shows test results. For the moment, we only consider the first row in Fig. \ref{fig:ONtest_smallest_largest_diag_elem}, which corresponds to polar shapelets. The results agree with Fig.\ \ref{fig:det_XtX_shapelets}, revealing undersampling effects for small $\beta$ and boundary truncation for large $\beta$ \citep[cf.][and discussion therein]{Melchior2007}. Furthermore, we can see that the axis ratio has only a mild impact on the orthonormality. However, inspecting the other rows corresponding to steeper profiles ($n_S=1,2,4$), Fig. \ref{fig:ONtest_smallest_largest_diag_elem} reveals serious violations of orthonormality. Obviously, for $n_S\geq 1$ the undersampling regime is not overcome before the boundary effects set in. This is confirmed by the right-most column in Fig. \ref{fig:ONtest_smallest_largest_diag_elem}, which used an enlarged pixel grid while keeping the resolution (scale radius $\beta$) constant. In this case, s\'ersiclets with $n_S=1$ exhibit decent orthonormality on this larger grid, while the cases with $n_S>1$ still to not reach an acceptable level of orthonormality. The reason for this peculiar behaviour is the argument $b\left(r/\beta\right)^{1/n_S}$ of the associated Laguerre polynomials in Eq. (\ref{eq:radial_basis_fct}). It implies that the polynomials are only slowly varying with $r$ for large values of $n_S$. Consequently, polar s\'ersiclets have a serious problem with orthonormality, especially for large S\'ersic indices.

What is the impact of undersampling? If $\det(X^T\cdot X)$ is too close to zero, numerical inaccuracies will dominate its inversion and the estimate of the expansion coefficients (Eq.\ (\ref{eq:ML_estimate_coeffs})) will catch up random errors. Consequently, this increases the uncertainty in the model parameters and may even lead to completely random parameter values.

The impact of undersampling could be alleviated by sampling the model several times within each pixel. In the limit of infinitely fine sampling this amounts to a convolution of the model with the pixel-response function. As the undersampling problem  stems from a key feature of the employed s\'ersiclet model, namely rapid polynomial oscillations, an oversampled model may better approximate the image data. However, the information about the galaxy morphology contained in the expansion coefficients has been lost and is not restored by oversampling. Instead, excessive oversampling generates artificial information that compromises the model parameters. As we require meaningful information from the expansion coefficients, we will not further pursue this approach. If one wants to employ this approach, one has to bear in mind, that the additional convolution will lead to covariances among the coefficients and therefore render the fitting procedure more complicated.

\subsection{Analogy to Nyquist frequency\label{sect:solution_1}}

In order to avoid the undersampling problem in practice, e.g., in morphological classification, we want to derive a lower limit to the scale radius $\beta$ of a s\'ersiclet model. This limit is given by comparing the pixel size to the scale on which the radial components of s\'ersiclets -- essentially the Laguerre polynomials -- vary. In fact, this is an analogy to the Nyquist frequency in the case of Fourier transform.

The key to set this lower limit is to identify the shortest scale on which a Laguerre polynomial varies. This scale can be inferred from the roots of the Laguerre polynomial. An associated Laguerre polynomial $L_l^k(x)$ of order $l$ has $l$ real-valued, positive roots within the interval $(0,l+k+(l-1)\sqrt{l+k}]$. The smallest scale that can be resolved by this polynomial is given by the distance between $x=0$ (the peak of the radial component) and the first root $x_1>0$. Unfortunately, the roots of associated Laguerre polynomials are not known analytically, so the first root $x_1$ has to be inferred numerically. As the argument of the associated Laguerre polynomial is $x=b(r/\beta)^{1/n_S}$, given $x_1$, we can infer the radius of the first root to be $r_1=\beta(x_1/b)^{n_S}$. Hence, the distance between $r=0$ and the first root is given by
\begin{equation}
\Delta r=r_1-0=r_1=\beta(x_1/b)^{n_S} \;\textrm{.}
\end{equation}
This scale $\Delta r$ should be well resolved by the pixel grid in order to avoid undersampling. An optimistic lower limit is $\Delta r=1$, i.e., the radial component drops from its central peak to its first root within a single pixel. Setting $\Delta r=1$ and solving for the minimal $\beta$, we obtain,
\begin{equation}\label{eq:lower_limit_beta}
\beta\geq \beta_\textrm{min}= (b/x_1)^{n_S}\;\textrm{,}
\end{equation}
where $\beta$ and hence $\beta_\textrm{min}$ are given in units of pixels. Table~\ref{tab:lower_limit_beta} provides values of $\beta_\textrm{min}$ for some realistic values of $n_S$ and maximum polynomial order $\nmax=l$. Considering this table, we also have to keep in mind that the image grid has to be large enough such that several scale radii fit into it. Evidently, as $n_S$ increases, the limit becomes larger and larger. This can also be inferred qualitatively from Fig. \ref{fig:ONtest_smallest_largest_diag_elem}, but Eq. (\ref{eq:lower_limit_beta}) provides a quantitative result. Obviously, for $n_S\geq 2$ the lower limit becomes extraordinarily large. In particular, $n_S=4$ would require an extremely well resolved object with radius of several thousand pixels.

\begin{table}
\begin{tabular}{cccccc}
\hline\hline
$\nmax$ & $n_S=0.5$ & $n_S=1.0$ & $n_S=2.0$ & $n_S=3.0$ & $n_S=4.0$ \\
\hline
 1 & 0.8165 & 0.8333 & 0.8403 & 0.8424 & 0.8435 \\
 2 & 1.0668 & 1.3145 & 1.7599 & 2.2042 & 2.6658 \\
 3 & 0.3256 & 1.7810 & 2.9324 & 4.3311 & 6.0286 \\
 4 & 1.4377 & 2.2423 & 4.3625 & 7.3782 & 11.537 \\
 5 & 1.5904 & 2.7011 & 6.0513 & 11.499 & 19.888 \\
 6 & 1.7296 & 3.1586 & 7.9993 & 16.848 & 31.873 \\
 7 & 1.8584 & 3.6151 & 10.207 & 23.577 & 48.375 \\
 8 & 1.9787 & 4.0712 & 12.674 & 31.841 & 70.372 \\
 9 & 2.0921 & 4.5268 & 15.400 & 41.792 & 98.931 \\
10 & 2.1996 & 4.9821 & 18.386 & 53.584 & 135.22 \\
11 & 2.3021 & 5.4373 & 21.632 & 67.371 & 180.49 \\
12 & 2.4002 & 5.8922 & 25.137 & 83.305 & 236.08 \\
13 & 2.4944 & 6.3471 & 28.902 & 101.54 & 303.46 \\
14 & 2.5853 & 6.8018 & 32.926 & 122.23 & 384.13 \\
15 & 2.6730 & 7.2565 & 37.210 & 145.53 & 479.75 \\
16 & 2.7579 & 7.7110 & 41.754 & 171.59 & 592.02 \\
17 & 2.8403 & 8.1656 & 46.557 & 200.57 & 722.76 \\
18 & 2.9204 & 8.6201 & 51.620 & 232.61 & 873.87 \\
19 & 2.9983 & 9.0745 & 56.943 & 267.88 & 1047.4 \\
20 & 3.0742 & 9.5289 & 62.525 & 306.52 & 1245.3 \\
\hline
\end{tabular}
\caption{Lower limits of scale radii $\beta$ in units of pixels for different maximum radial orders $\nmax$ according to Eq. (\ref{eq:lower_limit_beta}). Here we used $b=2n_S-1/3$.}
\label{tab:lower_limit_beta}
\end{table}

In the case of shapelets, it is standard practice to choose a maximum order $\nmax$ for a decomposition according to signal-to-noise ratio and resolution. For sersiclets, the resolution constraint plays a much more important role and restricts models with large $n_s$ to low $\nmax$, with which complicated morphological features could receive improper description. As elliptical galaxies rarely exhibit such features though, this restriction may not be overly problematic.

\section{Application to weak-lensing data}
\label{sect:app:lensing}

Given the fact that shapelets were also intended for shear measurements in weak-gravitational lensing \citep[e.g.,][]{Refregier2003b,Chang2004}, we now study the potential of s\'ersiclets in this field of research. From the simulated \textsc{Great08} data \citep{Bridle2009} we selected ten images with artificial galaxies, whose shear values are given in Table \ref{tab:Great08_sets}.\footnote{This restriction on only ten sets was necessary, because analysing the whole \textsc{Great08} data set would have been too time-consuming.} Every set contains 10,000 objects sampled on a 40$\times$40 pixel grid. Within such a set the applied shear is constant and the aim is to retrieve its value. Furthermore, all objects have been convolved with a PSF that is a Moffat profile with FWHM$=2.85$, $\beta = 3.5$, and intrinsic ellipticity $\epsilon=0.019 -0.007i$ \citep[cf.][]{Bridle2010}.

\begin{table}
\begin{tabular}{lrr}
\hline\hline
Set & $g_1$ & $g_2$ \\
\hline
0007 &  0.0005405 &  0.0069236\\
0026 & -0.0527875 & -0.0090224\\
0035 &  0.0166067 & -0.0045223\\
0048 &  0.0708862 & -0.0377040\\
0056 &  0.0325078 &  0.0978346\\
0091 & -0.0246346 & -0.0488837\\
0126 &  0.0170977 & -0.1383142\\
0135 &  0.0596913 &  0.0416342\\
0268 & -0.0653126 & -0.0883511\\
0281 & -0.0431769 &  0.0462176\\
\hline
\end{tabular}
\caption{Data sets chosen from the \textsc{Great08} sample and their complex shears $g=g_1+i g_2$.}
\label{tab:Great08_sets}
\end{table}

\begin{figure}
\includegraphics[width=8.1cm]{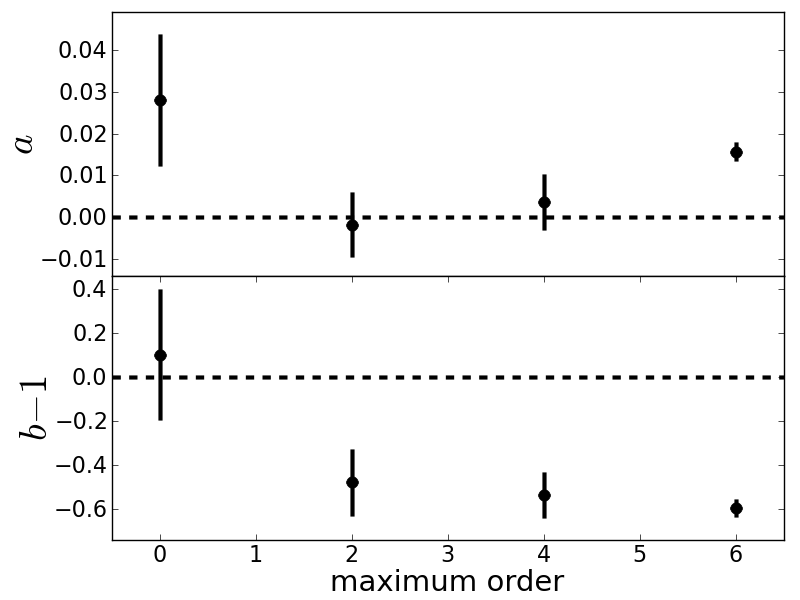}
\caption{Offset $a$ (top) and deviation from unity slope $b-1$ (bottom) of \textsc{Great08} data sets given in Table \ref{tab:Great08_sets} for different maximum orders, $\nmax$, of s\'ersiclet decomposition. Horizontal dashed lines indicate the ideal case of $a=0$ and $b=1$. Errors of $a$ and $b$ were estimated by bootstrap fitting of $f(x)=a+bx$ to the estimated and input shear values.}
\label{fig:Great08_biases}
\end{figure}

We decompose all objects into s\'ersiclets with maximum orders $\nmax=0,2,4,6$, taking into account the PSF by forward modelling. For each set, the shear is estimated via the mean ellipticity $\hat\epsilon$ from the 10,000 artificial galaxies via the ellipticity parameter of the s\'ersiclet model, which theoretically provides an unbiased estimator of the gravitational shear \citep[cf.][]{Bartelmann2001}. Figure \ref{fig:Great08_biases} shows the test results as input vs.\ estimated shear. The goodness of the shear estimates is parametrised by a straight-line model
\begin{equation}
f(x)=a+bx \;\textrm{.}
\end{equation}
A perfect shear estimator yields $a=0$ and $b=1$. If a real shear estimator yields an offset $a\neq 0$, i.e., a shear is detected although the input shear was zero, this typically implies that the PSF is not properly corrected for. If $b<1$, the true shears are underestimated.

Given the test results shown in Fig.\ \ref{fig:Great08_biases}, the s\'ersiclets perform best for $\nmax=0$, where the slope $b$ is consistent with 1 (at a fairly large offset $a$). This is not surprising, because for $\nmax=0$ s\'ersiclets reduce to pure S\'ersic profiles, which were used to simulate this subset of the \textsc{Great08} data.\footnote{The data sets were taken from the Great08 RealNoiseKnown branch, for which the galaxies are modeled as either of bulge or disc type, i.e. having S\'ersic index of $n_S=1$ or $n_S=4$, respectively.} However, the real lesson from Fig.~\ref{fig:Great08_biases} is that for $\nmax>0$ the slopes are significantly below~1, i.e., the shear is significantly underestimated. The higher-order s\'ersiclet modes, which should account for any deviation the actual galaxy shape exhibits with respect to the axisymmetric Sersic profile, do more harm than good as they become increasingly prone to undersampling. Low resolution is a generic feature of weak-lensing data, i.e., shear estimates based on s\'ersiclet decompositions always suffer from substantial undersampling effects. Oversampling of the model within each pixel could cure this, but is -- at least in our implementation -- computationally infeasible. Finally, we note that we observed a bias in the shear estimates based on Sersiclets. The first component of the shear is systematically overestimated whereas the second component is systematically underestimated. We do not yet understand the precise origin of this bias but we may speculate that this is a pixellation effect \citep[cf.\ Sect.~5.3 of][]{Massey2007}.

\section{Outlook: Orthonormalising higher-order Taylor expansions}
\label{sect:3rd_order_Taylor}

As discussed in Appendix~\ref{app:Sersic_and_Taylor}, the S\'ersic profile is the first-order Taylor expansion of any light profile. This naturally leads us to the expectation that with improving imaging quality the S\'ersic profile will not be a good match anymore, because it is ``only'' a \textit{first}-order expansion. Consequently, an obvious strategy to enhance the S\'ersic profile is to allow for higher orders in the Taylor expansion of Eq.\ (\ref{eq:Taylor_expansion}). Such higher-order radial profiles then can also be orthonormalised in order to describe azimuthal structures. This approach appears to be very promising for investigations of galaxy morphologies, e.g., in the context of classification. However, it introduces further nonlinear model parameters which renders it inappropriate for investigations of weak lensing where notoriously little data is available to constrain the model.

\subsection{Third-order profiles}

\begin{figure}
\includegraphics[width=8cm]{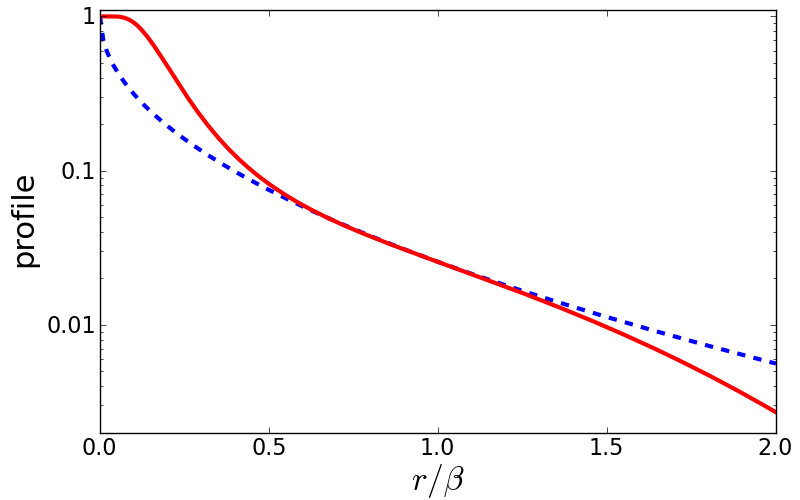}
\caption{Comparing a S\'ersic profile with $n_S=2$ (blue dashed line) to a third-order profile with $B=1/n_S=0.5$, $C=0.1$ and $D=0.25$ (red solid line).}
\label{fig:3rd_order_profile}
\end{figure}

As explained in Appendix~\ref{app:3rd_order_Taylor}, the next useful higher-order expansion which exhibits the correct boundary behaviour beyond the S\'ersic profile is a third-order expansion,
\begin{equation}\label{eq:3rd_order_profile}
p(r) \approx \exp\left[ -e^{A + B\log(r/\beta) + C\log^2(r/\beta) + D\log^3(r/\beta)}\right] \;\textrm{,}
\end{equation}
where $D>0$ and $A$, $B$, and $C$ are arbitrary. Figure \ref{fig:3rd_order_profile} shows an example of such a third-order profile in comparison to a normal S\'ersic profile. Evidently, the third-order profile can overcome two essential problems of pure S\'ersic profiles, since it (a) can exhibit a central cusp, which is also observed in real galaxies, and, (b) approaches zero for increasing radii faster than the pure S\'ersic profile. Therefore, we may speculate that such higher-order Taylor expansions could provide a reasonable generalisation, if deviations from the normal S\'ersic profile are observed while azimuthal structures are still absent. For instances, this may help to describe the light profiles of elliptical galaxies or unbarred S0 galaxies.

\begin{figure*}
\begin{center}
\includegraphics[width=16cm]{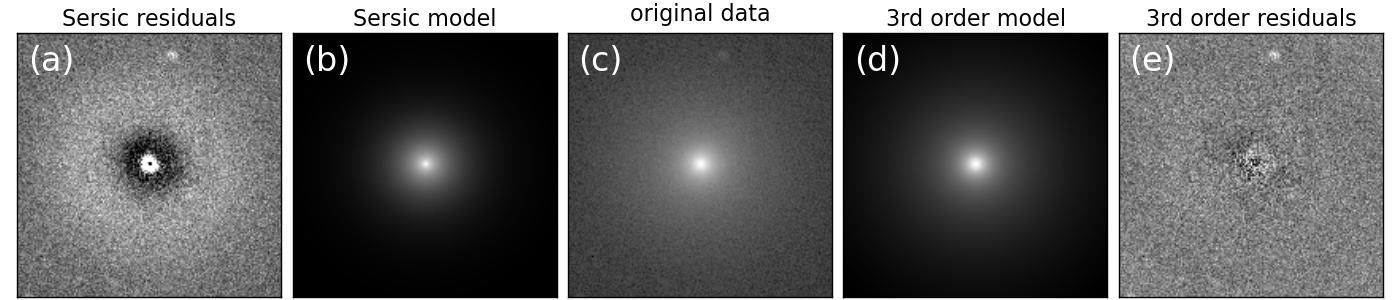}
\end{center}
\caption{Fitting S\'ersic and third-order profile to an elliptical galaxy. The panels denote residual map of S\'ersic fit (a), S\'ersic model (b), original data (c), best-fitting third-order profile (d), and residual map from third-order profile (e). Panels, b, c, and d have identical scaling. The residual maps (panels a and e) both use plot ranges from $-5\sigma$ to $5\sigma$.}
\label{fig:compare_fits_Sersic_3rdOrder_profiles}
\end{figure*}

\begin{figure}
\includegraphics[width=8cm]{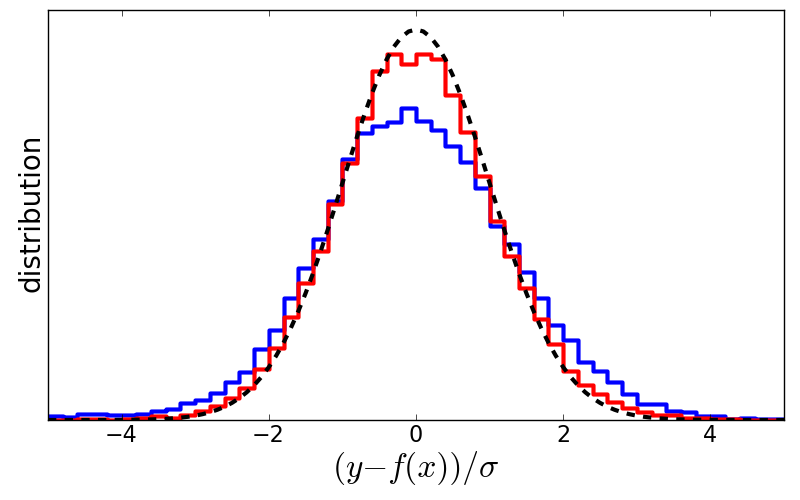}
\caption{Model comparison via distributions of normalised residuals. The blue histogram shows the distribution of normalised residuals from the S\'ersic fit. The red histogram are the normalised residuals from the third-order profile. The black dashed line is a unit Gaussian.}
\label{fig:normalised_residuals_Sersic_3rdOrder_profiles}
\end{figure}

In Fig.~\ref{fig:compare_fits_Sersic_3rdOrder_profiles} we compare the performances of the third-order profile and the S\'ersic profile by fitting an elliptical galaxy from the SDSS database. Clearly, the residual map of the third-order profile is almost perfectly random noise whereas the residual map of the S\'ersic profile reveals systematic mismodelling. Correspondingly, the ratio of $\chi^2$-values of the third-order profile over S\'ersic fit is $\chi^2_3/\chi^2_\textrm{S}\approx 0.48$. In fact, the third-order profile fits the data so well that we should ensure that it is not an overfit. For this purpose, Fig.~\ref{fig:normalised_residuals_Sersic_3rdOrder_profiles} displays the distributions of normalised residuals for both models in comparison to the unit Gaussian \citep[e.g.~see][]{Andrae2010d}. Evidently, the normalised residuals of the S\'ersic profile have a broader distribution than the unit Gaussian which is indicative of underfitting the data. The normalised residuals of the third-order profile are closer to the unit Gaussian, i.e., they are closer to the truth. However, their distribution does \textit{not} peak sharper than the unit Gaussian, i.e., the third-order profile does not overfit the data.

\subsection{Numerical orthonormalisation}

Third-order profiles such as Eq.~(\ref{eq:3rd_order_profile}) can be orthonormalised, too. Unfortunately, it is not possible to do this orthonormalisation analytically.\footnote{The reason is that the substitution in Appendix~\ref{app:derivation_sersiclets} is not bijective anymore.} Therefore, the orthonormalisation has to be performed numerically. We start with the radial monomials $(r^0,r^1,r^2,r^3,\ldots)$ which are linearly independent but not orthonormal. Using the definition of the scalar product according to Eq.~(\ref{app:eq:scalar_product}) and Dirac notation, we first normalise the ground state,
\begin{equation}
|0\rangle=\frac{|r^0\rangle}{\langle r^0|r^0\rangle} \;\textrm{.}
\end{equation}
Second, we compute the first-order state
\begin{equation}
|1\rangle=\left(\hat r-\langle 0|\hat r|0\rangle\right)|0\rangle \;\textrm{,}
\end{equation}
which is then also normalised such that $\langle 1|1\rangle=1$. Here, we used the following definition,
\begin{equation}
\langle k|\hat r|l\rangle = \int_0^\infty dr\,r\,B_k(r) r\,B_l(r) \;\textrm{,}
\end{equation}
where $B_l$ denotes the radial basis function of order $l$. The basis functions of order $l\geq2$ are then computed using the three-term recurrence relation
\begin{equation}
\qquad|l\rangle =\left(\hat r-\langle l-1|\hat r|l-1\rangle\right)|l-1\rangle
-\langle l-1|\hat r|l-2\rangle|l-2\rangle \,\textrm{.}
\end{equation}
This requires less numerical integrations than the brute-force Gram-Schmidt algorithm. Hence, it is faster and numerically more stable than the Gram-Schmidt algorithm.

\subsection{Revisiting the problems of s\'ersiclets}

The conceptual problem of s\'ersiclets was the postulated relation of steepness of the weight function and scale of polynomial oscillations that is not obeyed by real galaxies. This fixed relation is loosened by the third-order profiles. Of course, the S\'ersic index -- or rather $B=\frac{1}{n_S}$ -- still has an influence on the oscillation scale of the radial polynomials. However, now there are two further model parameters $C$ and $D$ which also influence the polynomials. Consequently, basis functions based on third-order profiles are more flexible and do not impose such a rigid relation as s\'ersiclets do.

\begin{figure}
\begin{center}
\includegraphics[width=8cm]{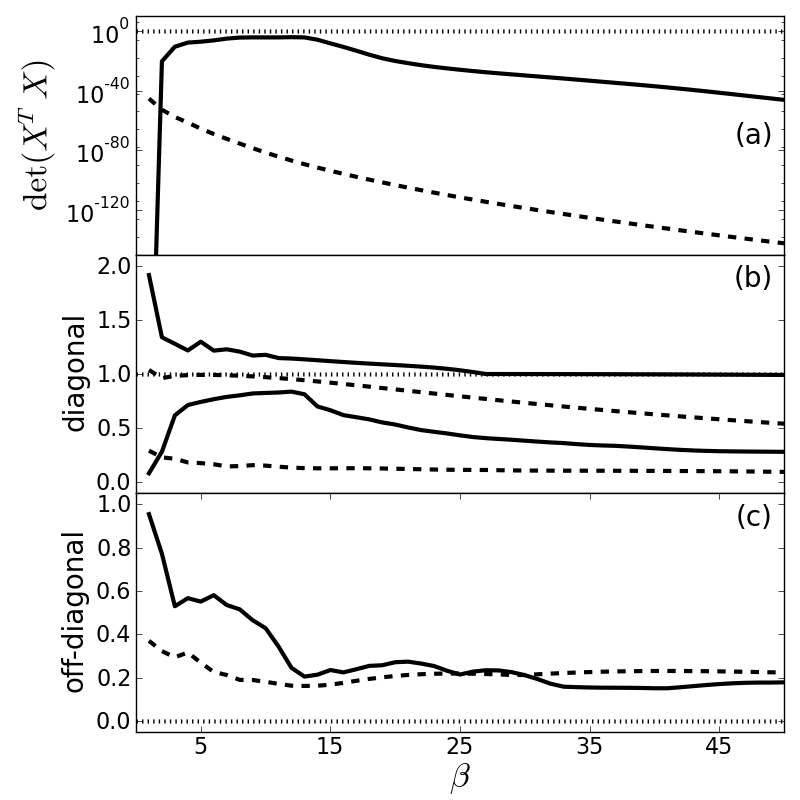}
\end{center}
\caption{Undersampling test for third-order profiles. We compare basis sets of the S\'ersic (dashed lines) and third-order profile (solid lines) shown in Fig.~\ref{fig:3rd_order_profile}. The test is performed on a 100$\times$100 pixel grid and the weight function has no intrinsic ellipticity. In contrast to Fig.~\ref{fig:ONtest_smallest_largest_diag_elem}, we have chosen $b=4-1/3$ in this case. Panel a: Determinant of coefficient covariance matrix of $X^T\cdot X$ which should be 1. Panel b: Largest and smallest diagonal elements of of $X^T\cdot X$ which should be 1. Panel c: Largest absolute value of off-diagonal elements of of $X^T\cdot X$ which should be 0.}
\label{fig:ON_test_3rd_order_profile}
\end{figure}

The practical problem of s\'ersiclets was undersampling, in particular for S\'ersic indices $n_S\geq 1$. In fact, Fig.~\ref{fig:3rd_order_profile} already suggests that third-order profiles can overcome this problem because they can exhibit central cusps instead of peaks. Therefore, the weights are not as highly concentrated and the polynomials may not oscillate as rapidly. We demonstrate this by repeating the orthonormality test of Fig.~\ref{fig:ONtest_smallest_largest_diag_elem} for the third-order profile with parameters given in Fig.~\ref{fig:3rd_order_profile} and its set of basis functions. Figure~\ref{fig:ON_test_3rd_order_profile} shows the test results. In direct comparison to s\'ersiclets with $n_S=\frac{1}{B}=2$, the orthonormalised third-order profile indeed suffers less strongly from undersampling, which is most obvious for the determinant of $X^T\cdot X$. In particular, there is a regime $5\leq\beta\leq 15$ where undersampling is overcome and boundary truncation has not yet set in. Such a regime does not exist for the corresponding set of s\'ersiclets shown here. Consequently, third-order profile can indeed overcome the undersampling problem of s\'ersiclets. Nevertheless, there are parameter choices for third-order profiles where the undersampling problem is as bad as for s\'ersiclets, e.g., when $C=D=0$ and the third-order profile reduces to a S\'ersic profile. However, by choosing appropriate priors it might be possible to exclude such parameter values in order to avoid undersampling.

\subsection{Computational feasibility}

We have shown that orthonormalisations of third-order profiles can overcome the limitations of s\'ersiclets while preserving their mathematical justification. However, we have not yet mentioned a serious practical limitation of this novel approach: The numerical orthonormalisation of third-order profiles is computationally highly expensive. Fitting a galaxy on a, say, 100$\times$100 pixel grid while freely adjusting all model parameters is infeasible on a standard computer. A problematic work-around could be to first fit a simple third-order profile to the galaxy and then only orthonormalise the best-fit profile in order to model the data's deviation from the mean profile. However, this step-wise approach by construction is very unlikely to find the best-fitting basis-function expansion because the first step produces a \textit{biased} fit and \textit{not} a mean profile. Therefore, we do not recommend this approach. Another work-around would be to set up a library of basis functions for all realistic parameter values of third-order profiles, such that numerical orthonormalisation has not to be conducted on-the fly during the fit. This approach is hampered by the fact that third-order profiles have three free parameters such that a decently detailed sampling of this three-dimensional parameter space -- which is necessary in order to provide reliable fits -- would produce an extensive library. The best solution is certainly to maintain the numerical orthonormalisation during the fit and to employ Graphics Processing Units (GPUs) instead of normal CPUs. For such a pure ``number crunching'' like numerical orthonormalisation, GPUs impressively outperform CPUs \citep[e.g.][]{Fluke2011}.

\section{Conclusions}
\label{sect:conclusions}

In this article, we reinvestigated the method of s\'ersiclets in order to overcome the limitations of shapelets reported by \citet{Melchior2009b} in parametrising galaxy morphologies. Orthogonalising the S\'ersic profile is justified by the fact that the S\'ersic profile is the first-order Taylor expansion of any real light profile (Appendix \ref{app:1st_order_Taylor}). We obtain the following results:
\begin{itemize}
\item From general radial light profiles like the S\'ersic profile, we can only construct polar basis function because Cartesian basis functions do not exhibit the azimuthal symmetry required in the context of galaxy morphologies. Shapelets are the only exception to this rule.
\item S\'ersiclet basis functions exist analytically, such that no numerical orthonormalisation is required and models are simple to evaluate. Polar shapelets are a special case of the s\'ersiclet basis functions.
\item S\'ersiclets outperform shapelets in the case of edge-on discs and elliptical galaxies, providing models with substantially lower residuals. In particular, s\'ersiclets overcome the ring-like artefacts introduced by shapelets when fitting galaxies with steep light profiles. However, in the case of extended objects with lots of substructure, e.g., face-on spiral galaxies, shapelets are competitive.
\item S\'ersiclets indeed solve the problems of shapelets. However, we revealed two new problems of this approach:
\begin{itemize}
\item S\'ersiclets are prone to undersampling effects, if galaxies are not well-resolved. Table~\ref{tab:lower_limit_beta} provides an estimate of the required resolution in order to avoid undersampling. This undersampling problem stems from the \textit{polar} basis functions being evaluated on a \textit{Cartesian} pixel grid. The problem is therefore most prominent in the central regions of the model, where unfortunately also most of the galactic light is concentrated. Undersampling gives rise to an increased uncertainty of the model parameters. For decreasing resolution, increasing parameter uncertainties render s\'ersiclet models less informative, which beyond some point obstructs most if not any application. If computationally feasible, the undersampling problem may be alleviated by oversampling the model, although oversampling inevitably compromises the interpretation of the expansion coefficients.
\item S\'ersiclets postulate a relation between steepness of the weight function and spatial scale of polynomial oscillations (cf.\ Sect.\ \ref{sect:interpret_nS}). However, real galaxy morphologies do not necessarily obey this relation, e.g., in the steepness of their bulge vs.\ the size and distribution of star-formation knots.\footnote{In fact, this problem also applies to shapelets.} Therefore, we have to expect modelling problems for s\'ersiclets, which may not have a generally predictable impact on a given application.
\end{itemize}
\item We tested the performance of s\'ersiclets by analysing simulated weak-lensing data from the \textsc{Great08} challenge \citep{Bridle2009}. We observed that higher-order modes do more harm than good. Given the typically low resolution of realistic weak-lensing images and the undersampling problem of s\'ersiclets, we have to conclude that s\'ersiclets are inappropriate for shear measurements in weak lensing.
\end{itemize}
We conclude that although the now formally correct s\'ersiclet approach appears very reasonable at first glance, it suffers from substantial problems in practice. Therefore, for every usage of s\'ersiclets, we recommend a critical assessment of the impact of undersampling and the postulated unphysical relation between steepness and oscillation scales onto the results.

Finally, we demonstrated that third-order Taylor expansions of galaxy light profiles can drastically improve the modelling fidelity in comparison to S\'ersic profiles. If orthonormlised, third-order profiles can overcome the undersampling problem of s\'ersiclets and loosen the tight relation between steepness of the weight function and spatial scale of polynomial oscillations. Furthermore, they preserve the mathematical justification of s\'ersiclets. However, such basis functions are computationally highly demanding. In fact, they may only be feasible using appropriate computer hardware such as GPUs.

\citet{JimenezTeja2011} introduce a set of basis functions based on Chebyshev rational functions. The authors demonstrate that their method is capable of providing excellent model reconstructions of galaxies of very different morphological types with very little computational cost. Orthonormalised higher-order profiles may provide similarly good models but certainly not at a similarly low computational cost. Hence, the method of \citet{JimenezTeja2011} appears to be superior.\footnote{Nevertheless, Chebyshev rational functions still have a long way to go. First, the choice of scale radius suggested by \citet{JimenezTeja2011} is a work-around that may erase discriminative information when it comes to classifying galaxy morphologies. Second, the authors yet have to investigate how their basis functions fare in the regime of poorly resolved galaxies which constitute the vast majority of galaxies in surveys.} In general, basis-function expansions can describe azimuthal structures of galaxy morphologies very accurately. However, the astrophysical interpretation of the expansion coefficients is not obvious and has to be laboriously deduced. This is a generic problem of basis-function expansions. Conversely, \textsc{Galfit3} \citep{Peng2010} allows to directly model the azimuthal structures exhibited by galaxy morphologies, such as spiral-arm patterns. The modelling and choice of components are ``guided'' by astrophysical intuition. Consequently, results obtained from \textsc{Galfit3} are easier to interpret. As discussed by \citet{JimenezTeja2011}, this requires substantial manual interaction of the user which is conceptually hard to automatise. The above mentioned methods highlight a fundamental conflict for any attempt of modelling galaxies. One can either seek to first optimally describe their shapes and then interpret the abstract model parameters later -- or attempt to immediately explain galaxy shapes as mixtures of a limited set of physically motived features.  The s\'ersiclets approach provides a compromise between these two extremes, as it incorporates the flexibility and completeness of basis-function approaches and the intuition of using a weight function, which is well-adapted to galactic morphologies.

The s\'ersiclet code is implemented in C++ and is available on request.

\section*{Acknowledgements}

We would like to thank the anonymous referee for valuable comments and suggestions that greatly improved the readibility of this paper. RA is funded by the Klaus Tschira Foundation via the Heidelberg Graduate School of Fundamental Physics (HGSFP). PM is supported by the U.S.\ Department of Energy under Contract No.\ DE-FG02-91ER40690. KJ is supported by the Emmy-Noether-programme of the DFG.

\bibliographystyle{aa}

\def\physrep{Phys. Rep.}%
\def\apjs{ApJS}%
\def\apj{ApJ}%
\def\aj{AJ}%
\def\aap{A\&A}%
\def\aaps{A\&AS}%
\def\mnras{MNRAS}%
\def\araa{ARA\&A}%
\def\pasa{PASA}%
\def\nat{Nature}%
\bibliography{bibliography}

\begin{thebibliography}{40}
\expandafter\ifx\csname natexlab\endcsname\relax\def\natexlab#1{#1}\fi

\bibitem[{{Abraham} {et~al.}(1996){Abraham}, {Tanvir}, {Santiago}, {Ellis},
  {Glazebrook}, \& {van den Bergh}}]{Abraham1996}
{Abraham}, R.~G., {Tanvir}, N.~R., {Santiago}, B.~X., {et~al.} 1996, \mnras,
  279, L47

\bibitem[{{Abraham} {et~al.}(2003){Abraham}, {van den Bergh}, \&
  {Nair}}]{Abraham2003}
{Abraham}, R.~G., {van den Bergh}, S., \& {Nair}, P. 2003, \apj, 588, 218

\bibitem[{{Andrae} \& {Jahnke}(in prep.)}]{Andrae2011}
{Andrae}, R. \& {Jahnke}, K. in prep.

\bibitem[{{Andrae} {et~al.}(2010{\natexlab{a}}){Andrae}, {Jahnke}, \&
  {Melchior}}]{Andrae2010b}
{Andrae}, R., {Jahnke}, K., \& {Melchior}, P. 2010{\natexlab{a}}, \mnras, 1836

\bibitem[{{Andrae} {et~al.}(2010{\natexlab{b}}){Andrae}, {Melchior}, \&
  {Bartelmann}}]{Andrae2010a}
{Andrae}, R., {Melchior}, P., \& {Bartelmann}, M. 2010{\natexlab{b}}, \aap,
  522, A21+

\bibitem[{{Andrae} {et~al.}(2010{\natexlab{c}}){Andrae}, {Schulze-Hartung}, \&
  {Melchior}}]{Andrae2010d}
{Andrae}, R., {Schulze-Hartung}, T., \& {Melchior}, P. 2010{\natexlab{c}},
  arXiv e-prints 1012.3754

\bibitem[{{Barlow}(1993)}]{Barlow1993}
{Barlow}, R. 1993, Statistics: A Guide to the Use of Statistical Methods in the
  Physical Sciences (Wiley VCH)

\bibitem[{{Bartelmann} \& {Schneider}(2001)}]{Bartelmann2001}
{Bartelmann}, M. \& {Schneider}, P. 2001, \physrep, 340, 291

\bibitem[{{Bernstein}(2010)}]{Bernstein2010}
{Bernstein}, G.~M. 2010, \mnras, 406, 2793

\bibitem[{{Bernstein} \& {Jarvis}(2002)}]{Bernstein2002}
{Bernstein}, G.~M. \& {Jarvis}, M. 2002, \aj, 123, 583

\bibitem[{{Berry} {et~al.}(2004){Berry}, {Hobson}, \& {Withington}}]{Berry2004}
{Berry}, R.~H., {Hobson}, M.~P., \& {Withington}, S. 2004, \mnras, 354, 199

\bibitem[{{Bershady} {et~al.}(2000){Bershady}, {Jangren}, \&
  {Conselice}}]{Bershady2000}
{Bershady}, M.~A., {Jangren}, A., \& {Conselice}, C.~J. 2000, \aj, 119, 2645

\bibitem[{{Bridle} {et~al.}(2010){Bridle}, {Balan}, {Bethge}, {Gentile},
  {Harmeling}, {Heymans}, {Hirsch}, {Hosseini}, {Jarvis}, {Kirk}, {Kitching},
  {Kuijken}, {Lewis}, {Paulin-Henriksson}, {Sch{\"o}lkopf}, {Velander},
  {Voigt}, {Witherick}, {Amara}, {Bernstein}, {Courbin}, {Gill}, {Heavens},
  {Mandelbaum}, {Massey}, {Moghaddam}, {Rassat}, {R{\'e}fr{\'e}gier}, {Rhodes},
  {Schrabback}, {Shawe-Taylor}, {Shmakova}, {van Waerbeke}, \&
  {Wittman}}]{Bridle2010}
{Bridle}, S., {Balan}, S.~T., {Bethge}, M., {et~al.} 2010, \mnras, 405, 2044

\bibitem[{{Bridle} {et~al.}(2009){Bridle}, {Shawe-Taylor}, {Amara},
  {Applegate}, {Balan}, {Bernstein}, {Dahle}, {Erben}, {Gill}, {Heavens},
  {Heymans}, {High}, {Hoekstra}, {Jarvis}, {Kirk}, {Kitching}, {Kneib},
  {Kuijken}, {Lagatutta}, {Mandelbaum}, {Massey}, {Mellier}, {Moghaddam},
  {Moudden}, {Nakajima}, {Paulin-Henriksson}, {Pires}, {Rassat}, {Refregier},
  {Rhodes}, {Schrabback}, {Semboloni}, {Shmakova}, {van Waerbeke}, {Witherick},
  {Voigt}, \& {Wittman}}]{Bridle2009}
{Bridle}, S., {Shawe-Taylor}, J., {Amara}, A., {et~al.} 2009, Annals of Applied
  Statistics, 3, 6

\bibitem[{{Chang} {et~al.}(2004){Chang}, {Refregier}, \& {Helfand}}]{Chang2004}
{Chang}, T., {Refregier}, A., \& {Helfand}, D.~J. 2004, \apj, 617, 794

\bibitem[{{Ciotti} \& {Bertin}(1999)}]{Ciotti1999}
{Ciotti}, L. \& {Bertin}, G. 1999, \aap, 352, 447

\bibitem[{{Conselice}(2003)}]{Conselice2003}
{Conselice}, C.~J. 2003, {The Relationship between Stellar Light Distributions
  of Galaxies and Their Formation Histories}

\bibitem[{{Ferry} {et~al.}(2008){Ferry}, {Rhodes}, {Massey}, {White}, {Coe}, \&
  {Mobasher}}]{Ferry2008}
{Ferry}, M., {Rhodes}, J., {Massey}, R., {et~al.} 2008, Astroparticle Physics,
  30, 65

\bibitem[{{Fluke} {et~al.}(2011){Fluke}, {Barnes}, {Barsdell}, \&
  {Hassan}}]{Fluke2011}
{Fluke}, C.~J., {Barnes}, D.~G., {Barsdell}, B.~R., \& {Hassan}, A.~H. 2011,
  \pasa, 28, 15

\bibitem[{{Graham} \& {Driver}(2005)}]{Graham2005}
{Graham}, A.~W. \& {Driver}, S.~P. 2005, Publications of the Astronomical
  Society of Australia, 22, 118

\bibitem[{{Jim{\'e}nez-Teja} \& {Ben{\'{\i}}tez}(2011)}]{JimenezTeja2011}
{Jim{\'e}nez-Teja}, Y. \& {Ben{\'{\i}}tez}, N. 2011, ArXiv e-prints

\bibitem[{{Kelly} \& {McKay}(2004)}]{Kelly2004}
{Kelly}, B.~C. \& {McKay}, T.~A. 2004, \aj, 127, 625

\bibitem[{{Kelly} \& {McKay}(2005)}]{Kelly2005}
{Kelly}, B.~C. \& {McKay}, T.~A. 2005, \aj, 129, 1287

\bibitem[{{Kennicutt}(1998)}]{Kennicutt1998}
{Kennicutt}, Jr., R.~C. 1998, \araa, 36, 189

\bibitem[{{Lotz} {et~al.}(2008){Lotz}, {Davis}, {Faber}, {Guhathakurta},
  {Gwyn}, {Huang}, {Koo}, {Le Floc'h}, {Lin}, {Newman}, {Noeske}, {Papovich},
  {Willmer}, {Coil}, {Conselice}, {Cooper}, {Hopkins}, {Metevier}, {Primack},
  {Rieke}, \& {Weiner}}]{Lotz2008}
{Lotz}, J.~M., {Davis}, M., {Faber}, S.~M., {et~al.} 2008, \apj, 672, 177

\bibitem[{{Lotz} {et~al.}(2004){Lotz}, {Primack}, \& {Madau}}]{Lotz2004}
{Lotz}, J.~M., {Primack}, J., \& {Madau}, P. 2004, \aj, 128, 163

\bibitem[{{Massey} \& {R\'efr\'egier}(2005)}]{Massey2005}
{Massey}, R. \& {R\'efr\'egier}, A. 2005, \mnras, 363, 197

\bibitem[{{Massey} {et~al.}(2007){Massey}, {Rowe}, {Refregier}, {Bacon}, \&
  {Berg{\'e}}}]{Massey2007}
{Massey}, R., {Rowe}, B., {Refregier}, A., {Bacon}, D.~J., \& {Berg{\'e}}, J.
  2007, \mnras, 380, 229

\bibitem[{{Melchior} {et~al.}(2010){Melchior}, {B{\"o}hnert}, {Lombardi}, \&
  {Bartelmann}}]{Melchior2009b}
{Melchior}, P., {B{\"o}hnert}, A., {Lombardi}, M., \& {Bartelmann}, M. 2010,
  \aap, 510, A75+

\bibitem[{{Melchior} {et~al.}(2007){Melchior}, {Meneghetti}, \&
  {Bartelmann}}]{Melchior2007}
{Melchior}, P., {Meneghetti}, M., \& {Bartelmann}, M. 2007, \aap, 463, 1215

\bibitem[{Nelder \& Mead(1965)}]{Nelder1965}
Nelder, J.~A. \& Mead, R. 1965, The Computer Journal, 7, 308

\bibitem[{{Ngan} {et~al.}(2009){Ngan}, {van Waerbeke}, {Mahdavi}, {Heymans}, \&
  {Hoekstra}}]{Ngan2008}
{Ngan}, W., {van Waerbeke}, L., {Mahdavi}, A., {Heymans}, C., \& {Hoekstra}, H.
  2009, \mnras, 396, 1211

\bibitem[{{Peng} {et~al.}(2010){Peng}, {Ho}, {Impey}, \& {Rix}}]{Peng2010}
{Peng}, C.~Y., {Ho}, L.~C., {Impey}, C.~D., \& {Rix}, H. 2010, \aj, 139, 2097

\bibitem[{{R\'efr\'egier}(2003)}]{Refregier2003}
{R\'efr\'egier}, A. 2003, \mnras, 338, 35

\bibitem[{{R\'efr\'egier} \& {Bacon}(2003)}]{Refregier2003b}
{R\'efr\'egier}, A. \& {Bacon}, D. 2003, \mnras, 338, 48

\bibitem[{{Rosa-Gonz{\'a}lez} {et~al.}(2002){Rosa-Gonz{\'a}lez}, {Terlevich},
  \& {Terlevich}}]{Rosa2002}
{Rosa-Gonz{\'a}lez}, D., {Terlevich}, E., \& {Terlevich}, R. 2002, \mnras, 332,
  283

\bibitem[{{S\'ersic}(1968)}]{Sersic1968}
{S\'ersic}, J.~L. 1968, {Atlas de galaxias australes} (Cordoba, Argentina:
  Observatorio Astronomico, 1968)

\bibitem[{{Simmat} {et~al.}(2010){Simmat}, {Tuffs}, \& {Popescu}}]{Simmat2010}
{Simmat}, E., {Tuffs}, R.~J., \& {Popescu}, C.~C. 2010, in American Institute
  of Physics Conference Series, Vol. 1240, American Institute of Physics
  Conference Series, ed. {V.~P.~Debattista \& C.~C.~Popescu}, 87--88

\bibitem[{{Slosar} {et~al.}(2009){Slosar}, {Land}, {Bamford}, {Lintott},
  {Andreescu}, {Murray}, {Nichol}, {Raddick}, {Schawinski}, {Szalay}, {Thomas},
  \& {Vandenberg}}]{Slosar2009}
{Slosar}, A., {Land}, K., {Bamford}, S., {et~al.} 2009, \mnras, 392, 1225

\bibitem[{{van der Wel} {et~al.}(2010){van der Wel}, {Bell}, {Holden},
  {Skibba}, \& {Rix}}]{Wel2010}
{van der Wel}, A., {Bell}, E.~F., {Holden}, B.~P., {Skibba}, R.~A., \& {Rix},
  H. 2010, \apj, 714, 1779

\end{thebibliography}

\appendix

\section{S\'ersic profile as first-order Taylor expansion}
\label{app:Sersic_and_Taylor}

We reveal the nature of the S\'ersic profile as a first-order Taylor expansion. Our emphasis is on the justification of the choice of the S\'ersic profile for orthogonalisation.\footnote{\citet{Ciotti1999} also investigate a Taylor-expansion of the S\'ersic profile, but they expand in powers of S\'ersic index and \textit{not} in powers of radius.} Furthermore, we briefly outline an alternative approach to generalise S\'ersic profiles by higher-order expansions.

\subsection{First-order expansion}
\label{app:1st_order_Taylor}

Let us consider the real, two-dimensional light profile of a galaxy $I(\vec x)$ projected onto the sky, which may exhibit arbitrary radial and azimuthal structures. Due to observational effects, e.g., noise and resolution, the observed light profile $I_{obs}(\vec x)$ will be different. If noise and resolution have a strong impact, we cannot identify azimuthal structures anymore and only the radial decline is left, i.e., $I_{obs}(\vec x)\approx I_{obs}(r)$. Let us further consider a rescaling of the observed light profile $p(r)=I_{obs}(r)/I_{obs}(0)$, such that $0<p(r)\leq 1$. Then we can take the logarithm of $p(r)$ and due to $\log p(r) \leq 0$, for $r>0$ we can also take the logarithm a second time, introducing
\begin{equation}
\tilde p(r) = \log(-\log p(r)) \;\textrm{.}
\end{equation}
We now Taylor expand this function $\tilde p(r)$ in $\log r$ at a characteristic radius $\log\beta$ to first order, i.e., a constant plus the first nontrivial term, which yields,
\begin{equation}\label{eq:Taylor_expansion}
\tilde p(r) \approx A + B(\log r-\log\beta)=A+\log(r/\beta)^B \;\textrm{.}
\end{equation}
Let us transform backwards now,
\begin{equation}
\log p(r) = -e^{\tilde p(r)} \approx -e^A (r/\beta)^B
\end{equation}
and hence
\begin{equation}
p(r) \approx \exp\left[-e^A (r/\beta)^B\right] \;\textrm{.}
\end{equation}
All we need to do now is to identify the coefficients $A$ and $B$ of the Taylor expansion in Eq.\ (\ref{eq:Taylor_expansion}). If we simply ``rename'' these constants by $B=1/n_S$ and $A=\log(b_n)$, then we have arrived at the definition of the S\'ersic profile
\begin{equation}\label{eq:Sersic_profile}
p_S(r) = \exp\left[-b_n\left(\frac{r}{\beta}\right)^{1/n_S}\right] \;\textrm{.}
\end{equation}
Evidently, the S\'ersic profile is a first-order Taylor expansion at $r_0=\beta>0$. In other words, in the limit of low resolution and low signal-to-noise ratios (where substructures are negligible) \textit{any} radial light profile is approximately a S\'ersic profile.\footnote{This is not true for radial profiles that are not differentiable, i.e., where no Taylor expansion exists. However, such profiles are unphysical.} This naturally explains why the S\'ersic profile is such a good fit to real galaxies in this regime. Furthermore, this also leads us to the expectation that with improving imaging quality the S\'ersic profile will not be a good match anymore, because it is ``only'' a \textit{first}-order expansion.

\subsection{Higher-order expansions}
\label{app:3rd_order_Taylor}

Given the fact that the S\'ersic profile is the first-order Taylor expansion of a real light profile, an obvious strategy to enhance the S\'ersic profile is to allow for higher orders in the Taylor expansion of Eq.\ (\ref{eq:Taylor_expansion}). However, a realistic profile has to be unity at $r=0$ and it has to approach zero for $r\rightarrow\infty$. As the Taylor expansion is in $\log r$, only expansions where the leading order term is of odd power in $\log r$ and has a positive expansion coefficient can satisfy these constraints. Therefore, the next useful higher-order expansion beyond the S\'ersic profile is a third-order expansion,
\begin{equation}
\tilde p(r) \approx A + B\log(r/\beta) + C\log^2(r/\beta) + D\log^3(r/\beta) \;\textrm{,}
\end{equation}
where $D>0$ and $A$, $B$, and $C$ are arbitrary.

\section{Mathematical derivation of radial parts of s\'ersiclets}
\label{app:derivation_sersiclets}

In this appendix, we give the derivation of the radial part of the s\'ersiclets, showing that the basis functions have an analytic form. Our starting point is the scalar product of two radial modes of orders $l$ and $l^\prime$,
\begin{equation}\label{app:eq:scalar_product}
\langle l|l^\prime \rangle = \int_0^\infty dr\,r\,R_l(r/\beta) R_{l^\prime}(r/\beta)\exp\left[-b\left(\frac{r}{\beta}\right)^{1/n_S}\right]
\end{equation}
where we have adopted Dirac notation from quantum mechanics. $R_l(r)$ denotes the radial polynomials we are looking for and the S\'ersic profile acts as the weight function or ``metric'' of this scalar product. Note the functional determinant $dr\,r$, which is due to our integration in polar coordinates. We now change variables according to
\begin{equation}\label{app:eq:substitution}
u(r)=b\left(\frac{r}{\beta}\right)^{1/n_S} \,\textrm{,}
\end{equation}
such that
\begin{equation}
dr\,r = \beta^2 \frac{n_S}{b^{2 n_S}} u^{2 n_S - 1} du \,\textrm{.}
\end{equation}
The limits of integration do not change under this transformation. Then Eq. (\ref{app:eq:scalar_product}) reads
\begin{equation}
\langle l|l^\prime \rangle = \beta^2 \frac{n_S}{b^{2 n_S}} \int_0^\infty du\,\tilde R_l(u) \tilde R_{l^\prime}(u)u^{2 n_S - 1}e^{-u} \,\textrm{.}
\end{equation}
The ``new'' weight function of this transformed scalar product is now of the form $u^k e^{-u}$, where $k=2 n_S - 1$, and the corresponding set of orthogonal polynomials are the associated Laguerre polynomials\footnote{http://mathworld.wolfram.com/LaguerrePolynomial.html},
\begin{equation}\label{app:eq:associated_Laguerre}
L_l^k (u)=	\frac{e^u u^{-k}}{l!}\frac{d^l}{du^l}\left(e^{-u}u^{l+k}\right)
\,\textrm{.}
\end{equation}
The $L_l^k$ exist if and only if $k> -1$. This is guaranteed, since $k=2n_S-1$ and the S\'ersic index satisfies $n_S>0$. The normalisation factor is
\begin{equation}
\int_0^\infty du \,L_l^k (u)L_l^k (u) u^k e^{-u} = \frac{\Gamma(l+k+1)}{l!}
\,\textrm{,}
\end{equation}
where $\Gamma$ denotes the Gamma function. Consequently, the radial parts of the s\'ersiclets read
\begin{equation}\label{app:eq:radial_basis_function}
R_l(r) = \frac{1}{N_l} L_l^{2n_S-1} \left[b\left(r/\beta\right)^{1/n_S}\right] \exp\left[-\frac{b}{2}\left(r/\beta\right)^{1/n_S}\right]
\end{equation}
with normalisation factor
\begin{equation}\label{eq:app:normalisation_factor}
N_l = \sqrt{\frac{\beta^2 n_S}{b^{2 n_S}} \frac{\Gamma(l+2n_S)}{l!} } \,\textrm{.}
\end{equation}
Mind the factor of $1/2$ in the exponent of Eq. (\ref{app:eq:radial_basis_function}), which arises from our definition that the S\'ersic profile is the weight function in Eq. (\ref{app:eq:scalar_product}). We could also have defined Eq. (\ref{app:eq:radial_basis_function}) with a pure S\'ersic profile, resulting in a factor 2 arising in Eq. (\ref{app:eq:scalar_product}). Fit results obtained from both definitions will not differ, since these factors are absorbed in the parameter $b$.

In the special case of polar shapelets these expressions simplify. First, for $n_S=0.5$, we note that $k=2n_S-1=0$, which implies $\Gamma(l+k+1)=\Gamma(l+1)=l!$ and simplifies the associated Laguerre polynomial $L_l^k$ to the ``normal'' Laguerre polynomial $L_l^0=L_l$. Second, in order to obtain shapelets, we need to set $b=1$. Therefore, the radial parts of polar shapelets read
\begin{equation}\label{app:eq:radial_basis_function}
R_l(r) = \frac{1}{\sqrt{\beta^2/2}} L_l \left[\left(r/\beta\right)^2\right] \exp\left[-\frac{r^2}{2\beta^2}\right] \textrm{.}
\end{equation}
The basis functions introduced by \citet{Massey2005} as ``polar shapelets'' (their Eq.\ (8)) differ from Eq. (\ref{app:eq:radial_basis_function}) by using $x^{|m|}L_{(l-|m|)/2}^{|m|}(x^2)$ instead of $L_l(x^2)$. Their basis functions are not orthogonal, e.g., consider
\begin{equation}
\langle l=1,m=1|l=2,m=0\rangle\propto\int_0^\infty dr\,r\,e^{-r^2} r^{|1|} L_0^1(r^2)r^{|0|} L_1^0(r^2) \,\textrm{,}
\end{equation}
which equals $-\sqrt{\pi}/8$ and is thus non-zero. Consequently, the ``polar shapelets'' defined by \citet{Massey2005} are incorrect. However, the linear independence is preserved because no two states with identical values of $m$ (otherwise the azimuthal parts $e^{im\varphi}$ would differ) and different values of $l$ can have identical $x^{|m|}L_{(l-|m|)/2}^{|m|}(x^2)$.

\section{Optimisation procedure}
\label{sect:optimisation}

Having defined s\'ersiclet models, we also need to clarify how to fit such models to given imaging data.

\subsection{Maximum-likelihood solution}
\label{app:ML_Solution}

For the moment, we assume that we are given estimates of the nonlinear model parameters $(N_\textrm{max},n_S,\beta,\vec x_0,\epsilon)$. We define a vector $\vec c$ of free parameters that form the coefficients. Furthermore, we write the image as a vector $\vec y$ of size $N$, where $N$ denotes the number of pixels in the given image. Finally, we define the so-called $N\times P$ design matrix $X$, such that $X_{np}$ is the $p$-th basis function evaluated at the $n$-th image pixel (after forward convolution with the PSF). With these definitions, we define the residual vector
\begin{equation}
\vec R = \vec y - X\cdot\vec c \;\textrm{,}
\end{equation}
which is an vector of size $N$ itself. Here ``$\cdot$'' denotes matrix multiplication. As the pixel noise of modern imaging data is usually Gaussian in excellent approximation, we are allowed to define
\begin{equation}
\chi^2 = \vec R^T\cdot\Sigma^{-1}\cdot\vec R \;\textrm{,}
\end{equation}
where $\Sigma$ denotes the $N\times N$ pixel covariance matrix. In the case of uncorrelated Gaussian noise with constant variance $\sigma^2$ in all pixels, $\Sigma$ takes the simple form $\Sigma=\diag(\sigma^2,\ldots,\sigma^2)=\sigma^2 I$. It is straightforward to show that the maximum-likelihood estimate that minimises $\chi^2$ is
\begin{equation}\label{eq:ML_estimate_coeffs}
\hat{\vec c} = (X^T\cdot\Sigma^{-1}\cdot X)^{-1}\cdot X^T\cdot\Sigma^{-1}\cdot\vec y \;\textrm{.}
\end{equation}
The matrix $X^T\cdot\Sigma^{-1}\cdot X$ has a special meaning: It is the $P\times P$ covariance matrix of the coefficients $\vec c$, which would follow from a Fisher analysis. In this formalism, we get this information for free.

The advantage of expressing the maximum-likelihood solution in terms of matrix operations is that we can employ fast and efficient linear-algebra algorithms. In our case, the basis functions are implemented in C++ and we use wrapper classes to import the packages ATLAS\footnote{http://math-atlas.sourceforge.net} and LAPACK\footnote{http://www.netlib.org/lapack/}.

\subsection{Optimising the nonlinear parameters for given $\nmax$}

We now assume that we are given an esimate of the maximum order, $\nmax$, and need to get estimates of $(n_S,\beta,\vec x_0,\epsilon)$. These estimates are derived from a Simplex algorithm \citep{Nelder1965} that we incorporate from the GNU Scientific Library (GSL)\footnote{http://www.gnu.org/software/gsl/manual/}. The Simplex algorithm is an iterative optimisation algorithm that does not employ derivatives but only evaluations of $\chi^2$. We also employ priors in order to restrict the model parameters to reasonable values. These priors in detail ensure that:
\begin{itemize}
\item The S\'ersic index is in the range $0.1\leq n_S\leq 8$.
\item The scale radius satisfies $\beta>0$.
\item The object centroid $\vec x_0$ is within the pixel grid.
\item The complex ellipticity satisfies $|\epsilon|<1$.
\end{itemize}
Within these reasonable parameter ranges, we employ flat priors.

\subsection{Estimating the maximum order}

In the case of shapelets, the maximum order, $\nmax$, is estimated via a reduced $\chi^2$. However, demanding that $\chi^2$ equals the number of degrees of freedom is justified if \textit{and only if} the model is purely linear \citep[e.g.,][]{Barlow1993,Andrae2010d}. In the case of s\'ersiclets, this assumption is definitely violated, since s\'ersiclets contain many nonlinear fit parameters. We rather recommend to estimate $\nmax$ by comparing the normalised residuals for models of different maximum orders. The optimal $\nmax$ is then defined by the model whose normalised residuals are closest to a Gaussian with mean zero and variance one \citep{Andrae2010d}.

Concerning large samples of galaxy images, one can also decompose all objects using identical $\nmax$. This approach may be favourable because many techniques to analyse the resulting catalogue of models require that all coefficient vectors have the same dimensionality \citep[e.g. clustering analysis, cf.][]{Kelly2004,Kelly2005,Andrae2010a}.

\bsp

\label{lastpage}

\end{document}